# A Model for Aging under Deformation Field, Residual Stresses and Strains in Soft Glassy Materials

Yogesh M Joshi

Department of Chemical Engineering,

Indian Institute of Technology Kanpur,

Kanpur 208016. INDIA.

E-mail: joshi@iitk.ac.in

**Abstract**

A model is proposed that considers aging and rejuvenation in a soft glassy material as respectively a decrease and an increase in free energy. The aging term is weighted by inverse of characteristic relaxation time suggesting greater mobility of the constituents induce faster aging in a material. A dependence of relaxation time on free energy is proposed, which under quiescent conditions, leads to power law dependence of relaxation time on waiting time as observed experimentally. The model considers two cases namely, a constant modulus when aging is entropy controlled and a time dependent modulus. In the former and the latter cases the model has respectively two and three experimentally measurable parameters that are physically meaningful. Overall the model predicts how material undergoes aging and approaches rejuvenated state under application of deformation field. Particularly model proposes distinction between various kinds of rheological effects for different combinations of parameters. Interestingly, when relaxation time evolves stronger than linear, the model predicts various features observed in soft glassy materials such as thixotropic and constant yield stress, thixotropic shear banding, and presence of residual stress and strain.

# I. Introduction

Glassy soft materials such as concentrated suspensions and emulsions, foams, colloidal gels and variety of different pastes are routinely used in industry as well as in everyday life. In this class of materials either the crowding of constituting entities and/or inter-particle attractive/repulsive interactions kinetically restrict the same from achieving the equilibrium structures.[1-3] However, microscopic mobility of the constituents arising from the thermal energy induces slow but steady structural evolution to form progressively stable structures. This process of structural recovery is also known as physical aging,[3] wherein free energy of a material decreases as a function of time. If such material is subjected deformation field, the structure evolved during aging gets altered, which usually causes reversal of physical aging.[4] The corresponding process is termed as rejuvenation. The rheological behavior of soft glassy materials (SGMs) is determined by competition between aging and rejuvenation for a given deformation field, which leads to many unusual and sometimes opposite effects such as time dependent yield stress,[5-8] viscosity bifurcation,[9,10] shear banding,[5,11-14] delayed yielding,[15,16] delayed solidification,[17,18] overaging,[19-21] presence of residual stresses[22] and strains,[23,24] etc. In this paper we present a model that accounts for aging and rejuvenation in terms of evolution of free energy influenced by deformation field. In addition to describing many of the above mentioned experimental behaviors, the model prescribes criterion for their occurrence based on the behaviors under quiescent conditions.

In a process of physical aging relaxation time and sometimes elastic modulus of a glassy material evolve as a function of time while attaining progressively low free energy states.[24-29] As a result a solid-like character of a glassy material increases gradually as a function of time. Application of deformation field attenuates the rate of evolution of relaxation time and eventually causes decrease in relaxation time. In the limit of sufficiently strong deformation field, the time evolution of material stops and material (shear) melts to form a liquid.[15,24,27] Subsequent to the shear melting the physical aging reinitiates in a material. In a traditional rheology literature this phenomenon is represented as thixotropy.[30] SGMs also demonstrate yield stress; and depending upon whether yield stress evolves with time or remains constant, the materials have been respectively termed as thixotropic and simple yield stress materials.[5] While the recent literature indeed proposes existence of real yield stress in both thixotropic and simple yield stress materials, it has been long argued in the rheology literature that existence of real yield stress is a myth and in reality material only undergoes transition from a weak flowing regime to a strong flowing regime leading to so called engineering yield stress.[7]

The SGMs have also been observed to demonstrate shear banding [31]. In the thixotropic yield stress materials, constitutionally the stress does not exist for the strain rates below the critical value.[5, 13] Consequently, imposition of strain rate below the critical value leads to banding, wherein one band flows with the critical strain rate while the other does not flow. The relative width of each band depends on the values of imposed and critical strain rates. Existence of thixotropic yield stress also leads to viscosity bifurcation wherein application of stress below the threshold value cannot stop divergence of viscosity.[9] On the other hand, application of stress above the threshold leads to viscosity to achieve a finite value as a function of time. Rather than showing viscosity bifurcation, some materials show delayed solidification or delayed yielding. In the former, application of stress, no matter how large it is, leads to either constant viscosity or decrease in viscosity for a prolonged period before showing sudden enhancement.[17] In delayed yielding, on the other hand, application of stress cannot restrict enhancement in viscosity as a function of time in the initial period. However, in the limit of long times, material undergoes sudden yielding thereby inducing the fluidity.[15, 16]

Under application of strong deformation field a material rejuvenates, and consequently material is in a liquid state. The aging of a material subsequent to rejuvenation can be monitored by applying no stress or constant strain. In the former case of no stress, strain recovers as a function of time. Interestingly, however, if a material is subjected to creep during the period of strain recovery, resultant strain may show a non-monotonic dependence on time, causing an apparent paradox as observed experimentally.[23, 24, 32] Instead, if strain is kept constant subsequent to the rejuvenation, stress relaxes. However depending on the characteristic feature of an SGM, stress may show a complete relaxation, power law dependence on time or a non-zero plateau (residual stress) in the limit of long times.[22] The effect of aging on both the phenomena, stress relaxation as well as strain recovery has, however, not ben studied theoretically.

Various models that capture rheological behavior of the thixotropic materials have been proposed in the rheology literature.[33] According to Mewis and Wagner,[30] there are three aspects common in such modeling approaches. The first one is evolution equation of empirical structure parameter (usually represented by $\lambda$), which indicates the instantaneous state of a material. The second aspect is a relationship between $\lambda$ and the rheological properties; while the third aspect is a constitutive equation that relates stress, strain and their derivatives through the rheological properties. Evolution equation of $\lambda$ essentially contains two terms: a buildup term and a destruction term representing aging and rejuvenation respectively. A comprehensive list of various expressions representing build up and destruction terms along with the constitutive equations have been reported in the literature.[30, 33] Coussot proposed that the models in this class can be represented by a simple evolution expression for an arbitrary structure parameter $\lambda$, given by:[34]

$$\frac{d\lambda}{dt} = \frac{1}{T_0} - Q(\lambda)\dot{\gamma}. \tag{1}$$

This expression suggests that the structure builds up with a constant timescale $T_0$, while the destruction term is proportional to strain rate $\dot{\gamma}$ with a prefactor $Q$ that grows with $\lambda$. Coussot and coworkers[9] showed that steady state stress and strain rate shows a non-monotonic relation for a suitable choice $Q(\lambda)$ and viscosity $(\eta(\lambda))$. A class of models has also been proposed by representing $\lambda$ as a fluidity that is as an inverse of characteristic relaxation time.[35, 36] By considering various functional forms for decrease in fluidity as a function of time (aging) and increase in the same as a function of deformation field (rejuvenation), Derec et al.[35] and Picard et al.[36] proposed different kinds of relationships between steady state stress and strain rate, including non-monotonic, which lead to variety of rheological phenomena shown by SGMs. Particularly the non-monotonic relation between steady state stress and strain rate leads to qualitative prediction of various important rheological behaviors reported for SGMs such as viscosity bifurcation, thixotropic yield stress and shear banding.

While thixotropy/fluidity models tend to capture essence of the physics associated with soft glassy dynamics, more rigorous models such as soft glassy rheology (SGR) model and mode coupling theory (MCT) have been developed to study the soft glassy dynamics. MCT is developed, in principle, for colloidal glasses wherein cage diffusion is known to get progressively sluggish as particle concentration increases.[37] MCT considers that since the cages are nothing but the surrounding particles, whose diffusion is also similarly affected, there exists a forward feedback mechanism that impedes relaxation of the fluctuations in density. Consequently at certain concentration the relaxation time diverges causing glass transition. MCT predicts an onset of glass transition well, and has been modified to include the effect of deformation field.[6] The present versions of MCT, however, do not demonstrate any physical aging. SGR model,[38] on the other hand, is primarily based on aging dynamics considered in Bouchaud's trap model.[39] SGR model divides a material in mesoscopic domains and tracks evolution of each as a function of time for a given deformation field. The effect of deformation field in SGR model is considered through strain and is modeled as an activated process. The relaxation time of an individual mesoscopic element directly depends on strain as: $\tau = \tau_0 \exp\left(\left[E - \frac{1}{2}k\gamma^2\right]/x\right)$, where $E$ is the depth of energy well in which an element is trapped, $\tau_0$ is an inverse of attempt frequency and $\frac{1}{2}k\gamma^2$ is energy gained by the element due to strain $\gamma$. Noise temperature $x$ suggests energy available for activation, and in a normalized form $x=1$ is a point of glass transition below which material shows physical aging. Upon cage diffusion element gets trapped in a new cage whose depth is obtained from a prior distribution. For a given deformation field and at any point in time, distribution of energy well depths, in which elements are trapped, is related to stress that gives the constitutive equation. Both MCT and SGR model demonstrate many experimentally observed rheological behaviors of SGMs;[1] and although mathematically and computationally demanding, these models render microscopic insight into the glassy dynamics intercepted by the deformation field.

Physical aging takes place not just in SGMs but also in polymer glasses, wherein enthalpy decreases as a function of time.[40, 41] Aging in polymer glasses is usually modelled by considering decrease in specific enthalpy to be a first order process.[41-43] Typically the departure from equilibrium is defined as: $\delta_h = h - h_\infty$, where $h$ is the specific enthalpy at any instance, while $h_\infty$ represents specific enthalpy at equilibrium. Under isothermal conditions, Kovacs, Aklonis, Hutchinson and Ramos (KAHR) in their seminal contribution proposed that:[42, 44, 45]

$$\frac{d\delta_h}{dt} = -\frac{\delta_h}{\tau(\delta_h)}, \tag{2}$$

where $\tau$ is relaxation time that depends on departure from equilibrium $\delta_h$. If $\tau$ is small, the time taken to establish equilibrium is also small. The dependence: $\tau = \tau(\delta_h)$ is obtained from Adam-Gibbs Theory and is given by[46, 47]

$$\tau = B \exp\left(C/Ts_c\right), \tag{3}$$

where $B$ and $C$ are constants, $T$ is temperature while $s_c$ is configurational entropy, which can be obtained by knowing the difference in heat capacity of material in crystal and liquid state. Interestingly this simple model, which considers aging to be a first order process, allows excellent prediction of the time dependent physical behavior of variety of amorphous polymers at different temperatures and upon step up and down temperature jumps.[45, 47] As material ages $s_c$ decreases, which causes increase in $\tau$. As a result, decrease in $\delta_h$ becomes increasingly sluggish as aging progresses. Similar to specific enthalpy, specific volume of a glassy material decreases upon aging. Consequently equivalent model has been developed by KAHR[45] by expressing departure from equilibrium in terms of specific volume and replacing equation (3) by the empirical relation proposed by Doolittle,[48] which relates relaxation time to the free volume.

## II. Model

The SGMs are thermodynamically out of equilibrium materials. Every material, which is not at thermodynamic equilibrium, has a natural tendency to approach the thermodynamic equilibrium state.[49] However in order to facilitate such approach, the microscopic constituents of the SGMs are needed to be sufficiently mobile (thermal energy). Typically the soft materials are exposed to constant $P$ (pressure) and constant (controlled) $T$ conditions. In addition, by virtue of incompressible nature of the same, these materials also do not undergo any change in $v$ (specific volume) as a function of time. Under such conditions, the equilibrium state in these materials can be characterized by minimization of either Gibbs ($g$) or Helmholtz free energy ($a$).[49] Since $g = a + Pv$, when $P$ and $v$ are constants, minimization of $g$ and $a$ are equivalent. Therefore, in the analysis below we discuss this scenario, only in terms of free energy. In the process of aging, under quiescent conditions, structure of an arrested soft material undergoes spontaneous evolution such that it progressively attains lower free energy as a function of time.

Typically in SGMs solid to liquid transition occurs upon application of a strong deformation field, a process typically known as rejuvenation or shear melting. The completely shear melted samples, immediately after the shear melting is stopped, can be considered to possess the highest free energy: $g_0$. On the other hand, the minimum value of free energy is associated with that of the thermodynamic equilibrium state and is given by: $g_\infty$. If decrease in free energy ($g$) with respect to time is assumed to be a first order process, we get:

$$\frac{d\phi}{d\tilde{t}} = -\frac{\phi}{\tilde{\tau}(\phi)}, \tag{4}$$

where $\phi$ is the normalized excess free energy defined as: $\phi = (g - g_\infty)/(g_0 - g_\infty)$. Furthermore, $\tilde{t} = t/\tau_0$ is dimensionless time and $\tilde{\tau} = \tau/\tau_0$ is dimensionless relaxation time, where $\tau_0$ is the relaxation time of a soft glassy material in its completely rejuvenated state $(\phi = 1)$. In equation (4), we assume that the rate of change in free energy is proportional to excess free energy divided by the time scale of structural rearrangement $[\tilde{\tau}(\phi)]$ in a material. This time scale is equivalent to the relaxation time of a material, which is suggestive of the mobility of the constituents in a material at any given $\phi$. As mentioned before, any material which is out of thermodynamic equilibrium, aspires to achieve the thermodynamic equilibrium. However material can be driven out of thermodynamic equilibrium in a trivial sense by perturbing an equilibrium material to high energy states. The consequent response that establishes equilibrium is merely a transient and not a physical aging if the relaxation time is constant. As suggested by Fielding and coworkers,[38] for any process to qualify as physical aging, its relaxation time must increase during the time over which the relaxation takes place. Consequently, $\tilde{\tau}(\phi)$ must be a decreasing function of $\phi$. In the SGMs while physical aging indeed causes decrease in free energy as a function of time, we cannot associate any thermodynamically measurable variable with decrease in free energy. Furthermore, SGMs having variety of different microstructures demonstrate remarkably similar form of the dependence of relaxation time on aging time. It is therefore no surprise that no empirical or otherwise relation is available in the literature to relate a structure to free energy and in turn to the relaxation time in SGMs.

In particulate suspensions, increase in volume fraction $(\varphi)$ of the suspended particles, which curbs the mobility of the same, is also known to cause increase in relaxation time $(\tau)$. The corresponding relation between $\tau$ and $\varphi$ is due to Krieger and Dougherty,[50] which has been extensively used in the literature, and has been experimentally validated for variety of suspension systems.[37] Furthermore, the mode coupling theory (MCT), which predicts onset of glass transition in the colloidal glasses well, also employs identical functional form as that of Krieger and Dougherty.[37] In both the forms $\tau$ of suspension diverges according to a power law $\left(\tau \sim \left[1 - \left(\varphi/\varphi^*\right)\right]^{-B}\right)$ as $\varphi$ approaches a certain threshold $\varphi^*$ associated with random close packing. On the other hand, in aging glassy materials, under constant concentration of constituents, mobility decreases continuously due to decrease in free energy. In this work, we therefore propose a relation between relaxation time and free energy, which has equivalent functional form to that proposed by Krieger - Dougherty or MCT. In case of some SGMs, including suspension of particles with hard sphere interactions, the relaxation time may diverge for values of free energy above the minimum (nonzero values of $\phi$). If such value of free energy is denoted by $\phi^*$ (at which the constituents do not possess mobility to facilitate relaxation), a generic form of the proposed expression is given by:

$$\tilde{\tau} = \left(1 - \frac{f(\phi)}{f(\phi^*)}\right)^{-\beta}, \qquad (5)$$

where $\beta$ is a parameter. In this expression, we use $f(\phi)$ since exact relation between microstructure and $\phi$ is not known. However, $f(\phi)$ must obey following two constraints: (1) $f(\phi)$ must be a monotonically increasing function of $\phi$, and (2) in order to satisfy $\tilde{\tau} = 1$ at $\phi = 1$, $f(\phi)=0$ at $\phi = 1$. Equation (4) can be solved using equation (5) to yield:

$$\frac{d}{d\tilde{t}}\left[\tilde{\tau}^{(\beta-1)/\beta}\right] = \frac{(\beta-1)\phi}{f(\phi^*)}\frac{df}{d\phi}. \qquad (6)$$

For various values of $\beta$, and for any arbitrary functional form of $f(\phi)$ that satisfies above two conditions, $\tilde{\tau}$ is expected to show stronger than linear, weaker than linear or linear dependence on $\tilde{t}$ according to equation (6). Equation (4) suggests that when $\tilde{\tau}$ increases stronger than linear, $\tilde{\tau}$ must diverge before system reaches the equilibrium state ($\phi^* > 0$). On the other hand, for a linear or weaker relationship system must approach equilibrium state in the limit: $\tilde{t} \to \infty$. Equation (6), therefore, suggests that value of $\beta$ is directly related to the strength of evolution of $\tilde{\tau}$ as a function of time, which in turn controls $\phi^*$. Furthermore, the above discussion imposes another constraint on $f(\phi)$ that: (3) at $\phi = 0$, $f(\phi)$, $\phi^*$ and $\beta$ should assume such values that $\tilde{\tau} \to \infty$ in that limit (It is well known that for many SGMs including a suspension of concentrated monodispersed particles, the lowest free energy state is a crystal state for which relaxation time is $\infty$. Therefore, for all those materials wherein aging results in acquiring the lowest free energy state, $\tilde{\tau}$ diverges as equilibrium state is approached: $\tilde{\tau} \to \infty$ in the limit of $\phi = 0$). We propose following functional form that satisfies all the above three constraints given by:

$$f(\phi) = \ln \phi. \tag{7}$$

The proposed expression of relaxation time given by equations (5) and (7) can now be used to solve differential equation (6) to obtain dependence of $\tilde{\tau}$ on $\tilde{t}$ under quiescent conditions.

The initial condition to solve equation (6) can be represented as: $\phi = \phi_{sm}$ (or $\tilde{\tau} = \tilde{\tau}(\phi_{sm})$) at $\tilde{t} = 0$, that is the moment shear melting is stopped (In principle if shear melting tends to rejuvenate the material completely, $\phi_{sm} = 1$ (or $\tilde{\tau} = 1$); however as shown below such possibility exists only if shear melting is carried out at shear rates $\dot{\gamma} \to \infty$). Assuming $A = (1-\beta)/\ln \phi^*$, the solution of equation (6) for a mentioned initial condition is given by:

$$\tilde{\tau} = \left[\tilde{\tau}(\phi_{sm}) + A\tilde{t}\right]^{\mu}, \tag{8}$$

where $\mu = \beta/(\beta - 1)$. When material is shear melted by using strong flow field for which $\phi_{sm} \approx 1$, equation (8) can be further simplified in the limit of long times $(A\tilde{t} >> 1)$, to represent:

$$\tilde{\tau} \approx (A\tilde{t})^{\mu} \qquad \cdots \qquad A\tilde{t} >> 1. \tag{9}$$

In a dimensional form equation (9) is represented by: $\tau \approx A^\mu \tau_0 \left(t/\tau_0\right)^\mu$. Interestingly relaxation time of many glassy materials, which include soft, molecular and spin glasses, demonstrate power law dependence on time given by equation (9).[24, 28, 38, 41, 51, 52] It is therefore interesting to see the proposed relation between $\tilde{\tau}$ and $\phi$ given by equation (5) with an assumption of equation (7) leads to experimentally observed power law dependence. It should be noted that values of $\mu<1$ represents sub-aging, $\mu>1$ represents hyper-aging, while $\mu=1$ represents a full aging scenario.[1, 2] Equation (5) can be rewritten in terms of $\mu$ and $A$ as:

$$\tilde{\tau} = \left[1 - \frac{\ln \phi}{\ln \phi^*}\right]^{\frac{\mu}{(1-\mu)}} \quad \cdots \quad \mu > 1, \tag{10}$$

$$\tilde{\tau} = \left[1 + (\mu-1)A \ln \phi\right]^{\frac{\mu}{(1-\mu)}} \quad \cdots \quad \mu < 1, \text{ and} \tag{11}$$

$$\tilde{\tau} = \phi^{-A} \quad \cdots \quad \mu = 1. \tag{12}$$

Equation (12) is obtained by solving equation (10) or (11) in the limit of $\mu \to 1$. It can be seen that for hyper-aging $(\mu > 1)$, $\tilde{\tau} \to \infty$ as $\phi \to \phi^*$, where $\phi^*$ is given by:

$$\phi^* = \exp\left(1/A(1-\mu)\right) \ldots \text{ for } \ldots \mu > 1, \tag{13}$$

indicating divergence of relaxation time before the equilibrium state is reached $\left(\phi^* > 0\right)$. In a case of hyperaging, owing to lack of mobility $(\tau \to \infty)$, a material remains frozen in a high free energy state.

Among various power law dependences represented by equations (10) to (12), the linear dependence of relaxation time on waiting time $(\mu = 1)$ has important practical significance. Firstly the linear dependence is observed experimentally for a very broad class of SGMs in absence of the deformation field. Such dependence is also observed for molecular as well as spin glasses.[41, 52] In addition, from a scaling point of view it is often argued that in absence of any externally dominating time scale, which is a typical case in glassy materials, the only naturally available imposed time scale is waiting time, which is the time elapsed since either thermal quench (molecular glasses) or mechanical quench/shear melting (SGMs).[19] Consequently the relaxation time scales as waiting time. In the literature, however, various SGMs have been reported to show sub-aging $(\mu < 1)$ or hyper-aging $(\mu > 1)$ behaviors.[26, 29] Such behaviors can originate from imposition of another field on a material, which tends to increase or decrease the characteristic timescale of a material beyond that can be achieved by merely a physical aging process. In case the process of time dependent decrease in free energy is not entirely physical, but partly chemical, so that it is irreversible, material tends to show hyper-aging dynamics.[29, 53]

It is usually observed that, in an aging process, modulus of the glassy materials either remains constant or increases as a function of time. However, even in the latter cases, enhancement in modulus is usually not as spectacular as that of relaxation time. Scaling argument suggests that if $E$ is the average depth of the energy wells in which constituents of a soft glass are arrested, modulus can be represented as energy density: $G = cE/b^3$, where $b$ is the characteristic length-scale (such as average inter-particle distance or network length) associated with a material and $c$ is constant of proportionality.[46] Consequently if $E$ remains constant throughout the aging process, modulus of a material will remain constant even if relaxation time shows increase as per equation (9). Such possibility arises if the aging behavior of a system is purely entropic. Such scenario is observed for particulate colloidal glasses with hard sphere interactions, wherein energy is identical for all the states, and aging is controlled by maximization of entropy ($s$). Such case can also be equivalently represented by minimization of free energy as: $g = h - Ts$, as for entropic systems $h$ is constant throughout the aging process under isothermal and isobaric conditions. Therefore for purely entropy controlled aging systems modulus can be represented as:

$$\tilde{G} = 1, \tag{14}$$

where $\tilde{G} = G/G_0$ is dimensionless modulus and $G_0$ is the constant modulus.

For those materials, wherein constituents share energetic interactions with each other, mean energy well depth $E$ increases as a function of time. In a limit of either equilibrium state $(\phi \to 0)$ or high free energy 'frozen' state $(\phi \to \phi^*)$, $E$ saturates to a constant value $E^*$. Over the regime where $E$ increases as a function of time, we assume the mean relaxation time to have the Arrhenius dependence on $E$, given by: $\tau = \tau_m \exp(E/k_B T)$, where $\tau_m$ is the microscopic relaxation time.[38] However as $\phi \to \phi^*$ or $\phi \to 0$, relaxation time no longer obeys Arrhenius relationship, as even though $\tilde{\tau} \to \infty$, $E$ saturates to a finite value $E^*$. Such behavior is often observed for molecular glasses, wherein relaxation time dependence deviates from Arrhenius - to - MCT - to - Vogel Fulcher as glass transition is approached.[37] Consequently, in a limit where Arrhenius relation is obeyed (for $\phi > \phi^* \geq 0$), the dependence of modulus on relaxation is easily obtained as:

$$\tilde{G} = 1 - \frac{\ln \tilde{\tau}}{\ln \tilde{\tau}_m} \ldots \text{ for } \ldots \phi > \phi^*, \tag{15}$$

where $\tilde{\tau}_m = \tau_m/\tau_0$ (it should be noted that $\tilde{\tau}_m < 1$ as discussed below, while $\tilde{\tau} \geq 1$), $\tilde{G} = G/G_0$ is dimensionless modulus where $G_0$ is the modulus associated with the state: $\phi = 1$, and is given by: $G_0 = -(ck_B T/b^3)\ln \tilde{\tau}_m$. However, as frozen state is approached $(\phi \to \phi^*)$, modulus saturates to a finite value while $\tilde{\tau} \to \infty$.

Application of the deformation field increases $\phi$. We assume that the rate of increase of $\phi$ to be directly proportional to rate of strain $(\dot{\gamma}_V)$ associated with the viscous (dissipative) flow weighted by $1-\phi$. Here $\dot{\gamma}_V$ is the second invariant of the rate of strain tensor $\dot{\boldsymbol{\gamma}}_V$ associated with the viscous flow, given by: $\dot{\gamma}_V = \sqrt{(\dot{\boldsymbol{\gamma}}_V : \dot{\boldsymbol{\gamma}}_V^\dagger)/2}$.[54] Consequently, equation (4) can be modified for evolution under application of deformation field as:

$$\frac{d\phi}{d\tilde{t}} = -\frac{\phi}{\tilde{\tau}(\phi)} + (1-\phi)\tilde{\dot{\gamma}}_V, \tag{16}$$

where $\tilde{\dot{\gamma}}_V = \tau_0 \dot{\gamma}_V$ is strain rate in a dimensionless form. Equation (16) is the evolution equation for $\phi$ under application of deformation field. The strain rate associated with viscous flow can be directly related with stress tensor as: $\boldsymbol{\sigma} = \eta \dot{\boldsymbol{\gamma}}_V$. Viscosity $\eta = G\tau$ is a product of relaxation time and modulus, which can respectively represented by equations (10) to (12) and equation (14) or (15). For a simple shear flow field equation (16) can therefore be modified to:

$$\frac{d\phi}{d\tilde{t}} = -\frac{\phi}{\tilde{\tau}(\phi)} + (1-\phi)\left(\frac{\tilde{\sigma}}{\tilde{\tau}\tilde{G}}\right), \tag{17}$$

where $\tilde{\sigma} = \sigma/G_0$ is dimensionless shear stress.

Usually the soft glassy materials are viscoelastic in nature. We can, therefore, use a single mode Maxwell model, which is the simplest constitutive equation for a viscoelastic material. For a time dependent modulus and viscosity a single mode Maxwell model is given by:

$$\dot{\boldsymbol{\gamma}} = \dot{\boldsymbol{\gamma}}_V + \dot{\boldsymbol{\gamma}}_E = \frac{\boldsymbol{\sigma}}{\eta} + \frac{d}{dt}\left[\frac{\boldsymbol{\sigma}}{G}\right]. \tag{18}$$

Here $\boldsymbol{\sigma}$ is stress tensor and $G$ and $\eta$ are time dependent modulus and viscosity of a material respectively. In equation (18) the first and the second terms are respectively the viscous and the elastic contributions to the strain rate. It is important to note here that in equation (16) it is assumed that $\phi$ gets affected only by the viscous component of the strain rate. This is because energy associated with elastic strain remains stored in a material and therefore the corresponding rate does not cause rejuvenation. We also show in the next section that even though stress is applied on a material in one direction (positive) or applied stress is zero there could be strain rate in the spring ($\dot{\boldsymbol{\gamma}}_E$) in the opposite direction (negative) due to increase in modulus or due to recovery. In this case, although $\dot{\boldsymbol{\gamma}}_E$ has a negative sign (assuming $\boldsymbol{\sigma}$ to be positive or zero) its second invariant will always have a positive sign. However, physically such reverse strain rate cannot cause rejuvenation in a material further justifying usage of only the viscous component of the strain rate in equation (16).

Equation (16) in a rate controlled form or equation (17) in a stress controlled form is the proposed expression for evolution of $\phi$. On the other hand, equation (18) is the constitutive equation associated with the model. Furthermore, we assume that the relation: $\tilde{\tau} = \tilde{\tau}(\phi)$ represented by equations (10) to (12) is intrinsic in nature and is independent of the nature and the strength of a deformation field. Therefore, a deformation field affects the evolution of relaxation time only through its dependence on $\phi$. As discussed before, under quiescent conditions (no deformation field), $\tilde{\tau}$ of a material shows power law dependence on $\tilde{t}$ as observed experimentally. Under application of deformation field, however, $\phi$ is expected to decrease or increase leading to increase or decrease in $\tilde{\tau}$.

Interestingly evolution of $\phi$ expressed by equation (16) can be transformed to a generic functional form given by equation (1) proposed by Coussot.[34] Multiplying equation (16) by $\tilde{\tau}/\phi$ leads to equation (1) with $\lambda = \int (\tilde{\tau}/\phi) d\phi$ and $Q = \tilde{\tau}(1-\phi)/\phi$. However unlike various previous approaches, that employ arbitrary functional forms for $Q = Q(\lambda)$ and $\eta = \eta(\lambda)$, the present model only needs expression of $\tilde{\tau}$ given by equation (5), which has been derived from physical arguments and comply with the experimental observation under quiescent conditions. For systems whose modulus increases with $\tilde{t}$, the present model has three parameters in a dimensionless form that are physically meaningful. The first is rate of aging $\mu$, the second is constant $A$ (which is equal to $\left[(1-\mu)\ln\phi^*\right]^{-1}$), and the third is $\tau_m$. However, if modulus is constant the model needs only the first two parameters: $\mu$ and $A$, which are the characteristics features of SGM that depend upon microstructure of the same. Most importantly $\mu$ and $A$ can be estimated experimentally by knowing dependence of relaxation time on aging time and have following constraints: $\mu \geq 0$ and $A > 0$. Such dependence can be very easily obtained by carrying out creep or stress relaxation experiments at different aging times as discussed in the literature.[15, 24, 27, 51, 55] In the present model microscopic relaxation time $(\tau_m)$ determines rate at which modulus evolves with time. Equation (15) suggests that smaller the value of $\tau_m$ is, weaker is the evolution of $\tilde{G}$. In the limit of $\phi = 1$, if mean depth of the energy wells occupied by the constituents of SGM is $E_0$, an Arrhenius relation leads to relaxation time of that state as: $\tau_0 = \tau_m \exp(E_0/k_B T)$, which leads to: $\tilde{\tau}_m = \exp(-E_0/k_B T)$. Although $E_0$ is the shallowest mean energy depth possible for $\phi = 1$, it is always positive. Consequently, $\tilde{\tau}_m$ must vary in the limit: $0 < \tilde{\tau}_m < 1$. (It is important to note that even though as per equation (15) it appears that in the limit of $\tau_m = 0$ modulus remains constant, such limit exists only if there is no aging. This is because microscopic relaxation time $\tau_m$ is a unit time with which a material ages. Even for a material wherein aging is purely entropic, wherein modulus is constant, $\tau_m$ is nonzero. This is because in such case relaxation time does not depend on energy well depth.)

## III. Results

To begin with we shall discuss results associated with the steady state predictions. In the limit of steady state, since $\dot{\gamma}_E = 0$, equation (16) leads to expressions for steady state strain rate given by:

$$\tilde{\dot{\gamma}}_{ss} = \frac{\phi_{ss}}{\tilde{\tau}_{ss}(1 - \phi_{ss})}. \tag{19}$$

On the other hand, equation (17) leads to the expression for steady state shear stress:

$$\tilde{\sigma}_{ss} = \tilde{G}_{ss} \frac{\phi_{ss}}{(1 - \phi_{ss})}. \tag{20}$$

In both the expressions, subscript $ss$ represents the steady state values of the respective variables (including $\tilde{\tau}_{ss} = \tilde{\tau}(\phi_{ss})$ given by equation (10) to (12) and $\tilde{G}_{ss} = \tilde{G}_{ss}(\phi_{ss})$). As expected, the steady state relationship between $\tilde{\sigma}$ and $\tilde{\dot{\gamma}}$ is simply:

$$\tilde{\sigma}_{ss} = \tilde{\tau}_{ss} \tilde{G}_{ss} \tilde{\dot{\gamma}}_{ss}, \tag{21}$$

where the constant of proportionality is dimensionless viscosity $\tilde{\eta} = \tilde{\tau}_{ss}\tilde{G}_{ss}$. In figure 1(a) we plot $\tilde{\sigma}_{ss}$ as a function of $\tilde{\dot{\gamma}}_{ss}$ for materials that show enhancement in modulus as a function of time for different values of $\mu$ at $A=10$ and $\tilde{\tau}_m=0.1$. It can be seen that the dependence of $\tilde{\sigma}_{ss}$ on $\tilde{\dot{\gamma}}_{ss}$ is monotonic for $\mu=1$ over an explored region, however becomes non-monotonic with a presence of a minima for the higher values of $\mu$. The region where $\tilde{\sigma}_{ss}$ decreases with increase in $\tilde{\dot{\gamma}}_{ss}$ is known to be unstable.[56] In figure 1b we also plot $\tilde{\sigma}_{ss}$ with respect to $\phi_{ss}$ by solving equation (20) for $A=10$ and $\tilde{\tau}_m=0.1$ for different values of $\mu$, which also shows non-monotonic relationships except for $\mu=1$. In the inset of figure 1b we plot relation between $\tilde{\sigma}_{ss}$ and $\phi_{ss}$ for $\mu=2$ but different values of $A$ and $\tilde{\tau}_m$. It can be seen that with increase in $\mu$ and $A$, the curves shift to greater values of $\tilde{\sigma}_{ss}$ and also shift $\phi_c$ (the value of $\phi_{ss}$ associated with the minimum in $\tilde{\sigma}_{ss}$) and $\phi^*$ (according to equation (13)) to higher values. The inset also shows behavior of the steady state curve at two values of $\tilde{\tau}_m = 0.1$ and $0.001$. Increase in $\tilde{\tau}_m$ shifts the location of minima as well as the curve to the higher values of $\tilde{\sigma}_{ss}$. As apparent from equations (19) to (21), the qualitative dependence of $\tilde{\sigma}_{ss}$ on $\tilde{\dot{\gamma}}_{ss}$ is similar to that of between $\tilde{\sigma}_{ss}$ and $\phi_{ss}$ with the minimum in $\tilde{\sigma}_{ss}$ in the former relation coinciding with that of the latter.

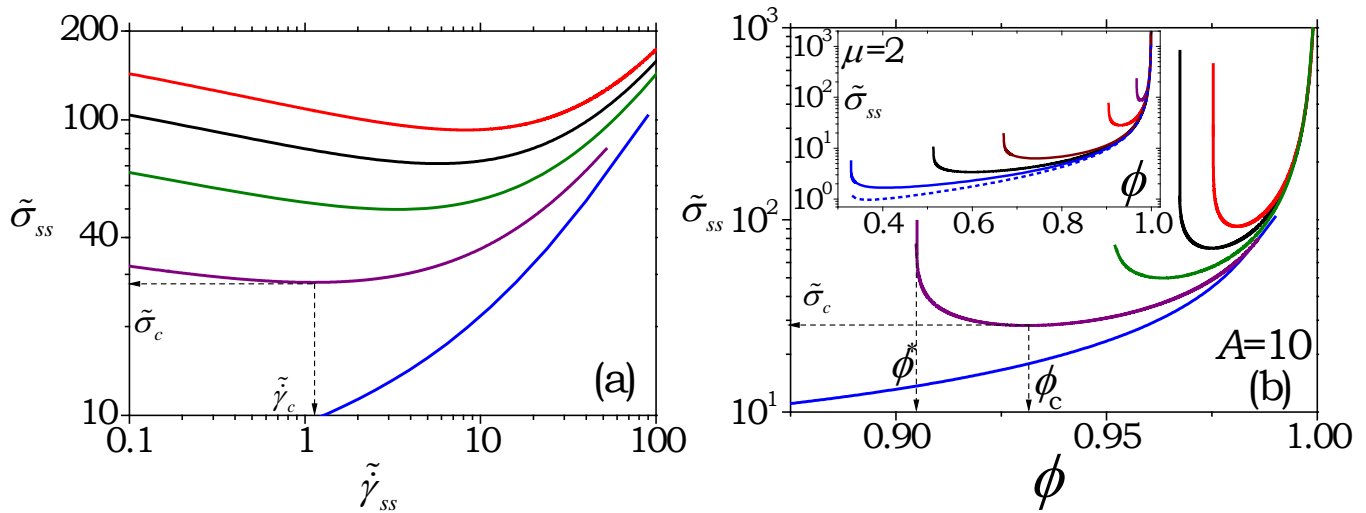

**Figure 1.** Relationship between $\tilde{\sigma}_{ss}$ and (a) $\tilde{\dot{\gamma}}_{ss}$, (b) $\phi$ given by equations (19) and (20) for different values of $\mu$ for $A=10$ and $\tilde{\tau}_m=0.1$. From bottom to top

$\mu = 1$, 2, 3, 4 and 5. In the inset of figure (b) $\tilde{\sigma}_{ss}$ is plotted against $\phi$ for $\mu = 2$ and different values of $A$ and $\tilde{\tau}_m$; while for a dotted line: $\tilde{\tau}_m = 0.001$ and $A = 0.9$. For full lines from top to bottom: $\tilde{\tau}_m = 0.1$ and $A = 30$, 10, 2.5, 1.5, and 0.9. In all the plots part of the curves having negative slope is an unstable region. The non-dimensional strain rate, stress and free energy associated with the minimum of the curve are represented by $\tilde{\dot{\gamma}}_c$, $\tilde{\sigma}_c$ and $\phi_c$ respectively.

In order to obtain the values of the parameters $\mu$, $A$ and $\tilde{\tau}_m$ for which flow curves become non-monotonic we solve $d\tilde{\sigma}_{ss}/d\tilde{\dot{\gamma}}_{ss} = 0$ by differentiating equation (21) by $\tilde{\dot{\gamma}}_{ss}$ leading to:

$$\frac{1}{1-\phi_c} + \left.\frac{d\ln\tilde{G}}{d\ln\phi}\right|_{\phi=\phi_c} = 0. \tag{22}$$

For a material with time dependent modulus, numerical solution of equation (22) gives $\phi_c$ from which $\tilde{\dot{\gamma}}_c$ and $\tilde{\sigma}_c$ (represented in figure 1) can be obtained by using equations (19) and (20) respectively for $\phi_{ss} = \phi_c$. In figures 2(a) and (b) we plot $\tilde{\dot{\gamma}}_c$, $\tilde{\sigma}_c$ and $\phi_c$ as a function of $\mu$ for various values of $A$ and $\tilde{\tau}_m$ for time dependent modulus given by equation (15). It can be seen that, irrespective of the values of $A$ and $\tilde{\tau}_m$, all the three variables: $\tilde{\dot{\gamma}}_c$, $\phi_c$ and $\tilde{\sigma}_c$ decrease with decrease in $\mu$; and tend to zero as $\mu$ approaches 1. Increase in $A$ as well as $\tilde{\tau}_m$, on the other hand, shifts all the curves to the higher values of respective ordinates. In figure 2(b) we also plot $\phi^*$, which is the minimum attainable value of $\phi$ (represented in figure 1) given by equation (13), with respect to $\mu$ for different values of $A$. There is no steady state associated with the values of $\phi$ in the range $\phi^* \leq \phi < \phi_c$ as it is an unstable branch. It can be seen that the width of the unstable region represented by $\phi_c - \phi^*$ decreases with increase in $A$ as well as $\mu$ (in the limit of $\mu \to 1$, both $\phi_c$ and $\phi^*$ approach zero). Furthermore equation (13) clearly shows that $\phi^*$ is independent of $\tilde{\tau}_m$. Figure 2(b) also shows that with decrease in $\tilde{\tau}_m$, $\phi_c$ decreases, and it can be shown from equations (15) and (22) that in the limit of $\tilde{\tau}_m \ll 1$, $\phi_c \to \phi^*$. Importantly figure 2 clearly indicates that the steady state stress – strain rate relationship is monotonic for $\mu \leq 1$.

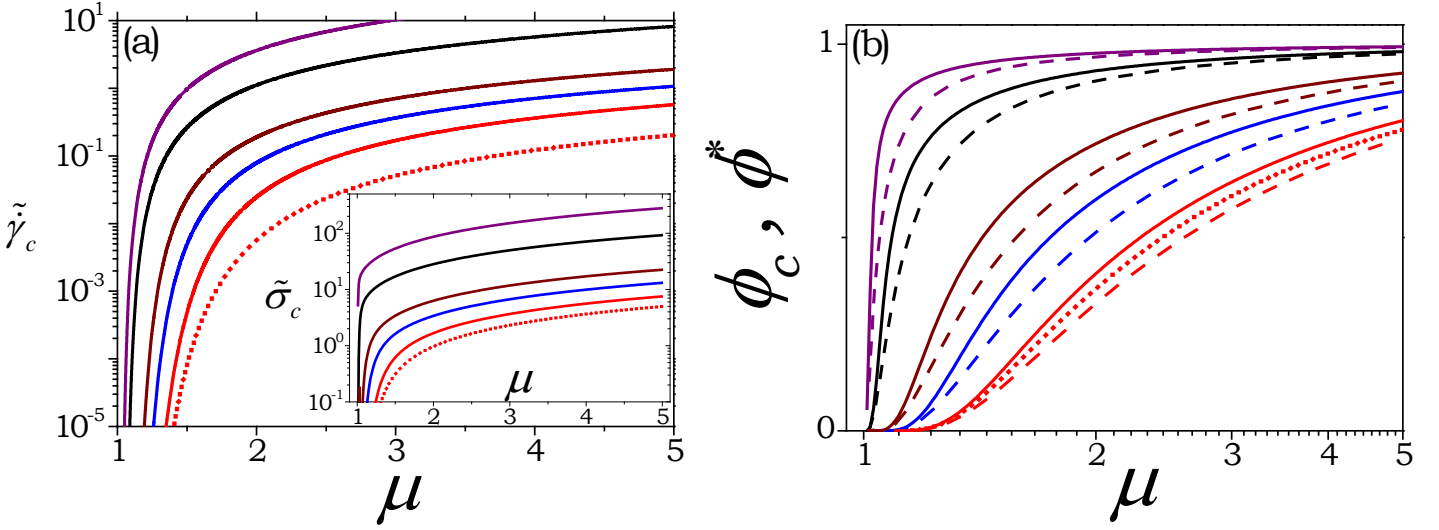

**Figure 2.** Dimensionless critical (a) strain rate $\left(\tilde{\dot{\gamma}}_c\right)$ and stress $\left(\tilde{\sigma}_c\right)$ (shown in inset) are plotted as a function of $\mu$. The full lines represent different values of $A$ (From top to bottom: 30, 10, 2.5, 1.5, and 0.9.) and $\tilde{\tau}_m = 0.1$. In figure (b) $\phi_c$ (full lines) and $\phi^*$ (dashed lines) [equation (13)] are plotted as a function of $\mu$. From top to bottom $A = 30$, 10, 2.5, 1.5, and 0.9. The dotted line in both the figures is for $A = 0.9$ and $\tilde{\tau}_m = 0.001$.

Now we consider a case when $\tilde{G} = 1$ during aging, for which equation (22) clearly indicates that dependence of $\tilde{\sigma}_{ss}$ on $\tilde{\dot{\gamma}}_{ss}$ does not show a minimum ($\phi_c$ does not exist in the range: $0 \leq \phi \leq 1$). Consequently for $\mu \leq 1$ dependence of $\tilde{\sigma}_{ss}$ on $\tilde{\dot{\gamma}}_{ss}$ must show a monotonic increase. For $\mu > 1$, according to equations (10) and (13), $\tilde{\tau} \to \infty$ as $\phi \to \phi^*$. As a result as $\tilde{\dot{\gamma}}_{ss} \to 0$ in the limit of $\tilde{\tau} \to \infty$, stress must show a plateau at:

$$\tilde{\sigma}_y = \frac{\phi^*}{1-\phi^*} = \frac{1}{\exp\left(1/A(\mu-1)\right)-1} \quad \ldots \quad \text{for } \tilde{G} = 1 \text{ and } \mu > 1, \quad (23)$$

where $\tilde{\sigma}_y$ is a yield stress. In figure 3 we plot $\tilde{\sigma}_{ss}$ as a function of $\tilde{\dot{\gamma}}_{ss}$ for different values of $\mu$ and $A$ at $\tilde{G}=1$. An observed plateau in $\tilde{\sigma}_{ss}$ in the limit of $\tilde{\dot{\gamma}}_{ss} \to 0$ indicates a presence of permanent yield stress that is independent of time (non-thixotropic yield stress). As shown in figure 3, $\tilde{\sigma}_y$ can be seen to be increasing with $\mu$ as well as $A$ as per equation (23).

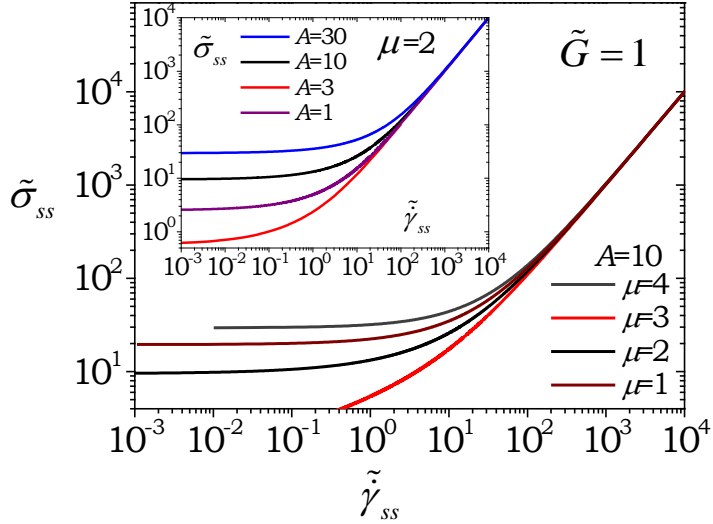

**Figure 3.** Relationship between $\tilde{\sigma}_{ss}$ and $\tilde{\dot{\gamma}}_{ss}$ given by equations (19) and (20) for different values of $\mu$ (From top to bottom $\mu=4, 3, 2$ and 1) for $A=10$ for a case when modulus remains constant $\tilde{G}=1$. It can be seen that except for $\mu>1$, $\tilde{\sigma}_{ss}$ shows a plateau in the limit of $\tilde{\dot{\gamma}}_{ss} \to 0$ demonstrating presence of constant yield stress. In the inset of $\tilde{\sigma}_{ss}$ is plotted against $\tilde{\dot{\gamma}}_{ss}$ for $\mu=2$ and different values of $A$ (From top to bottom, $A=30, 10, 3$ and 1). It can be seen that yields stress increases with $\mu$ and $A$ according to equation (23).

The presence of yield stress is also characterized by a non-monotonic flow curve, such as shown in figure 1, as there are no steady state values of strain rate ($\tilde{\dot{\gamma}}_{ss}$) associated with stresses smaller than that corresponding to the minimum represented by $\tilde{\sigma}_c$. This concept is described by figure 4, wherein we plot $\tilde{\sigma}_{ss}$ as a function of $\tilde{\dot{\gamma}}_{ss}$ for $A=10$, $\tilde{\tau}_m=0.1$ and two values of $\mu$: $\mu=1$ (Figure 4(a)) and $\mu=2$ (Figure 4(b)). We also plot the corresponding values of $\phi_{ss}$ on the abscissa. Let us consider a case, wherein subsequent to complete shear melting ($\phi=1$), a material is allowed to evolve without applying stress $\left(\tilde{\sigma}=0\right)$. Such evolution is carried out, wherein $\phi$ decreases as a function of time (according to equation (16) with $\tilde{\dot{\gamma}}_V=0$), until it reaches a certain value of $\phi=\phi_i$ (initial value of $\phi$) at which stress is applied. In figure 4(a), we consider a case wherein $\tilde{\sigma}=6$ is applied to a material. Consequently, if $\phi_i$ is in the region II, where $d\phi/d\tilde{t} < 0$, $\phi$ will continue to decrease until it reaches the steady state value associated with intersection of $\tilde{\sigma}_{ss}=6$ and the flow curve. If $\phi_i$ is in the region I, where $d\phi/d\tilde{t} > 0$, $\phi$ will increase until it reaches steady state value associated with $\tilde{\sigma}_{ss}=6$. However, since the flow curve is monotonic, a material will flow irrespective of the value of applied stress (The scenario for a material with constant modulus will be similar to that discussed for figure 4(a) as curves shown in figure 3 are also monotonic except the fact that those depict a plateau associated with permanent yield stress).

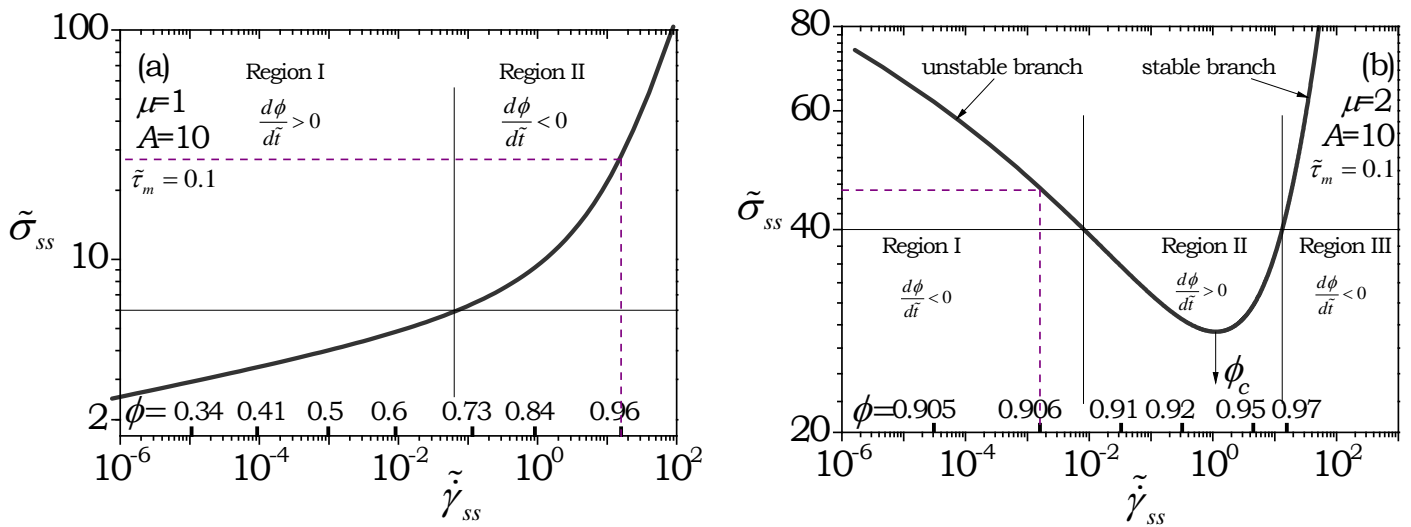

**Figure 4.** Steady state flow curve are shown for (a) $A=10$, $\tilde{\tau}_m=0.1$ and $\mu=1$ and (b) $A=10$, $\tilde{\tau}_m=0.1$ and $\mu=2$. The corresponding values of $\phi_{ss}$ are also shown on the inside part of an abscissa. For a monotonic flow curve a material will yield irrespective of the value of stress. For a non-monotonic flow curve, application of stress $\sigma_c$ on a material, will cause yielding (flow) only if $\phi_i > \phi_c$. In addition, if $\phi_i$ is in the range $\phi^* < \phi_i < \phi_c$, the application of stress will cause flow only if $d\phi/d\tilde{t}$ given by equation (17) is positive. Both the figures are discussed in detail in the text.

For figure 4(b), let us assume applied stress is $\tilde{\sigma}=40$. In this case the steady state value of $\phi$ is the one associated with intersection of $\tilde{\sigma}_{ss}=40$ and the increasing part of the flow curve. If $\phi_i$ is in region III, where $d\phi/d\tilde{t} < 0$, $\phi$ will continue to decrease until it reaches the steady state value. If $\phi_i$ is in region II, where $d\phi/d\tilde{t} > 0$, $\phi$ will increase until it reaches the steady state value. Therefore for a given applied stress greater than $\tilde{\sigma}_c$, if $\phi_i$ lies in regions II and III, a material will eventually attain a steady state. However if $\phi_i$ is in region I where $d\phi/d\tilde{t} < 0$, $\phi$ will continue to decrease even under application of the stress field until it attains the minimum possible value of $\phi^*$. Consequently a material will not attain the steady state.

The presence of non-monotonic flow curve as shown in figure 4(b), therefore leads to a natural dependence of yield stress on $\phi$ given by:

$$\tilde{\sigma}_y = \tilde{\sigma}_c \ldots \text{for} \ldots \phi_i \geq \phi_c \tag{24}$$

$$\tilde{\sigma}_y = \frac{\phi_i}{(1-\phi_i)} \frac{\ln\left[\tilde{\tau}(\phi_i)/\tilde{\tau}_m\right]}{\ln\left[1/\tilde{\tau}_m\right]} \ldots \text{for} \ldots \phi^* < \phi_i < \phi_c \tag{25}$$

Since $\phi_i$ decreases with time, the yield stress $\tilde{\sigma}_y$ will first remain constant for $\phi_i \geq \phi_c$ as shown by equation (24), and then increase with time for $\phi^* < \phi_i < \phi_c$ as per equation (25). In figure 5 we plot variation of $\tilde{\sigma}_y$ with $\tilde{t}$ for different values of $A$, $\tilde{\tau}_m$ and $\mu$. It can be seen that $\tilde{\sigma}_y$ is constant at small times and subsequently shows a logarithmic dependence on $\tilde{t}$. In addition, the dependence of $\tilde{\sigma}_y$ on $\tilde{t}$ becomes stronger with increase in all the three variables: $A$, $\tilde{\tau}_m$ and $\mu$. As explained in figure 4(b) and as described by equations (17) and (25), we can propose a thixotropic yielding criterion that upon application of stress $\sigma$ on a material in a momentary state $\phi_i$, if $\phi$ continues to decrease towards $\phi^*$ material will not yield. On the other hand, if application of stress causes evolution (increase or decrease) of $\phi$ so that it stabilizes at a value equal to or above $\phi_c$, a material will yield.

As described by equation (17), whether material yields or not, physical aging is affected by the strength of a stress field. Time evolution of relaxation time under a stress field can be obtained by manipulating equations (10) to (12) and (16), and is given by:

$$\mu_t = \frac{d \ln \tilde{\tau}}{d \ln \tilde{t}} = \mu \frac{A\tilde{t}}{\tau^{1/\mu}} \left[ 1 - \left\{ \frac{(1-\phi)}{\phi \tilde{G}} \right\} \tilde{\sigma} \right], \tag{26}$$

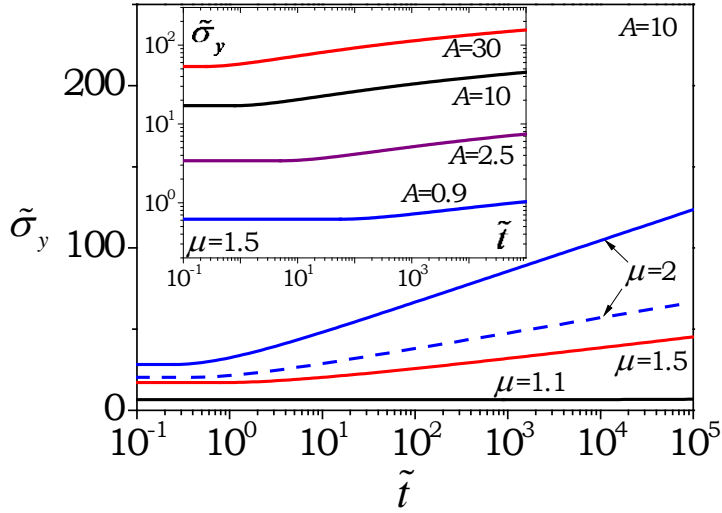

**Figure 5.** Evolution of dimensionless yield stress is plotted as a function of time for various values of $\mu$ for $A=10$ and $\tilde{\tau}_m=0.1$. The dashed line is for $\mu=2$ and $\tilde{\tau}_m=0.001$. Inset shows evolution of $\tilde{\sigma}_y$ for different values of $A$ at $\mu=1.5$ and $\tilde{\tau}_m=0.1$.

which clearly shows that for $\tilde{\sigma}=0$, relaxation time dependence described by equation (8) is recovered $\left(\mu_t = \mu\right)$. As discussed before, let us consider a case wherein material is allowed to age without applying stress, such that $\phi$ spontaneously decreases as per equation (4), and at $\phi = \phi_i$ stress is applied to a material. If $\phi_i \geq \phi_c$, the term in braces is simply reciprocal of $\tilde{\sigma}_{ss}\left(\phi_{ss} = \phi\right)$ (obtained by replacing $\phi_{ss}$ in equation (20) by $\phi$). Therefore, equation (26) can be expressed in a simpler format:

$$\mu_t = \frac{d\ln\tilde{\tau}}{d\ln\tilde{t}} = \mu\frac{A\tilde{t}}{\tilde{\tau}^{1/\mu}}\left[1 - \frac{\tilde{\sigma}}{\tilde{\sigma}_{ss}(\phi)}\right] \ldots \text{ for } \ldots \phi_i \geq \phi_c. \tag{27}$$

Consequently if $\phi_i \geq \phi_c$ and $\tilde{\sigma} > \tilde{\sigma}_c$, with increase in time $\tilde{\sigma}_{ss}$ tends to $\tilde{\sigma}$ so that $\mu_t$ must approach zero enabling a material to achieve the steady state. For the various values of parameters for which $\phi_c$ does not exist according to equation (22), evolution of relaxation time is given by either equation (26) or (27). We represent former case in figure 6(a) wherein we plot $\tilde{\tau}$ and $\mu_t$ as a function of time for $\mu = 1$, $\phi_i = 0.96$ (corresponding $\tilde{\sigma}_{ss} = 28.3$) for different values of $\tilde{\sigma}$. It can be seen that for $\tilde{\sigma} = 0$, $\tilde{\tau}$ shows continuous increase and corresponding $\mu_t$ approaches 1 in the limit of long times. Furthermore, for nonzero stresses, if $\tilde{\sigma} < \tilde{\sigma}_{ss}$ the evolution of $\tilde{\tau}$ weakens from the point of application of $\tilde{\sigma}$ leading to step decrease in $\mu_t$. The corresponding evolution of $\tilde{\tau}$, however, eventually plateaus out to a constant value causing $\mu_t$ to approach 0 after showing a maximum. If $\tilde{\sigma} > \tilde{\sigma}_{ss}$, $\tilde{\tau}$ decreases eventually leading to a plateau value and demonstrating negative values of $\mu_t$ before $\mu_t \to 0$. In figure 6(b) we also explore evolution of $\tilde{\tau}$ and $\mu_t$ for a system with constant modulus ($\tilde{G} = 1$), $\mu = 2$, and $\phi_i = 0.906$ (corresponding $\tilde{\sigma}_y = 9.5$, which is constant) for different values of $\tilde{\sigma}$ by solving equations (8) and (27). For $\tilde{\sigma} = 0$, evolution of $\tilde{\tau}$, as per equation (8) with $\phi_{sm} = 1$, attains $\mu_t = 2$ in the limit of long times. However for $\tilde{\sigma} < \tilde{\sigma}_y$, $\tilde{\tau}$ increases with time but with weaker dependence and the corresponding $\mu_t$ approaches $\mu$ in the limit of long times. Furthermore, since the flow curve for $\tilde{G} = 1$ is monotonic, for $\tilde{\sigma}_y < \tilde{\sigma} < \tilde{\sigma}_{ss}$ the behavior of $\tilde{\tau}$ and $\mu_t$ with respect to $\tilde{t}$ is expected to be qualitatively similar to that shown in figure 6(a) for $\tilde{\sigma} < \tilde{\sigma}_{ss}$. For $\tilde{\sigma} > \tilde{\sigma}_{ss} > \tilde{\sigma}_y$, $\mu_t$ continues to decrease and shows a minimum before approaching a steady state value of 0.

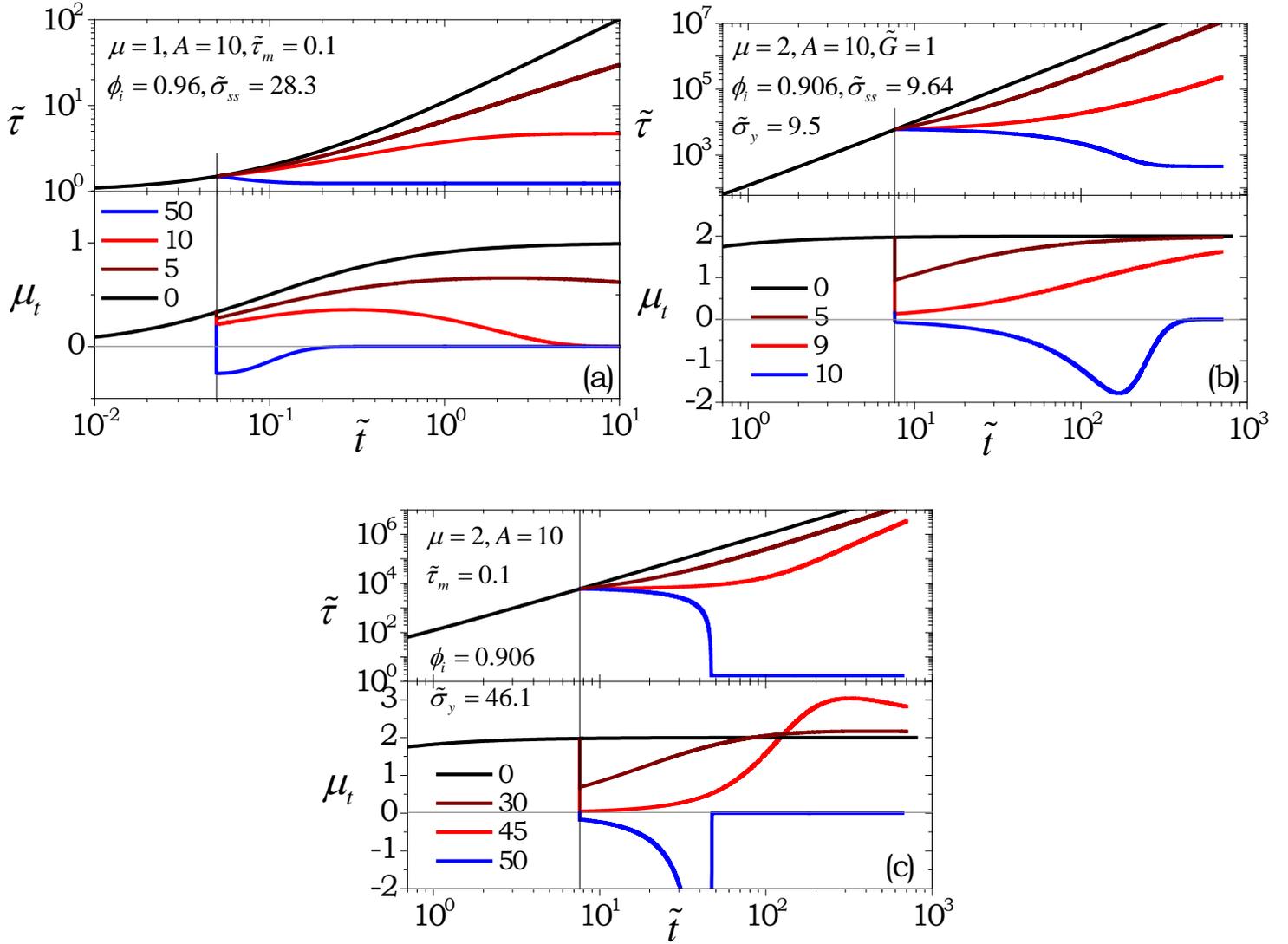

**Figure 6**. The temporal evolutions of $\tilde{\tau}$ and $\mu_t$ are plotted for different values of $\tilde{\sigma}$ for (a) $A=10$, $\tilde{\tau}_m=0.1$, and $\mu=1$ (The stress is applied when $\phi_i=0.96$ for which $\sigma_{ss}=28.3$). (b) The evolution of $\tilde{\tau}$ and $\mu_t$ are plotted for a system with $\tilde{G}=1$, $A=10$, $\mu=2$, and permanent yield stress $\tilde{\sigma}_y=9.5$ (The stress is applied when $\phi_i=0.906$). The values of stresses are shown in legend. In part (c) same variables are plotted for $A=10$, $\tilde{\tau}_m=0.1$ and $\mu=2$ (The stress is applied when $\phi_i=0.906$ for which $\tilde{\sigma}_y=46.1$). The corresponding positions of $\phi_i$ for (a) and (c) are described respectively in figure 3(a) and 3(b) by dotted lines.

For $\mu > 1$ and $\tilde{G}$ given by equation (15), the flow curve is non-monotonic. For such case if $\phi_i$ is such that: $\phi^* < \phi_i < \phi_c$, equation (25) suggests that the term in braces is essentially $\tilde{\sigma}_y(\phi)$. Consequently equation (26) can be rewritten as:

$$\mu_t = \frac{d \ln \tilde{\tau}}{d \ln \tilde{t}} = \mu \frac{A\tilde{t}}{\tau^{1/\mu}} \left[1 - \frac{\tilde{\sigma}}{\tilde{\sigma}_y(\phi)}\right] \ldots \text{for} \ldots \phi^* < \phi_i < \phi_c. \qquad (28)$$

We represent this scenario in figure 6(c) wherein time dependent evolution of $\tilde{\tau}$ and $\mu_t$ is plotted for $\mu=2$, $\phi_i=0.906$ (corresponding $\tilde{\sigma}_y=46.1$) for different values of $\tilde{\sigma}$. If $\tilde{\sigma} \geq \tilde{\sigma}_y(\phi_i)$, $\tilde{\sigma}_y(\phi) \to \tilde{\sigma}$ causing $\mu_t \to 0$ enabling material to attain the steady state. For $\tilde{\sigma} < \tilde{\sigma}_y(\phi_i)$, $\tilde{\tau}$ continues to increase but with weaker dependence. The corresponding $\mu_t$ shows a step decrease at the point of application stress, however increases subsequently. Very interestingly at moderately high times $\mu_t$ increases beyond $\mu=2$, and shows a maximum. Such behavior can be attributed to decrease in $\phi$ as a function of time which leads to $\tilde{\sigma}/\tilde{\sigma}_y(\phi) \to 0$ in the limit of long times. However owing to impeded increase in $\tau$ due to applied $\tilde{\sigma}$, $A\tilde{t}/\tau^{1/\mu}$ increases beyond unity causing $\mu_t$ to increase beyond $\mu$. Nonetheless as $\phi \to \phi^*$, $A\tilde{t}/\tau^{1/\mu}$ again decreases gradually.

Presence of yield stress in thixotropic materials ($\mu > 1$), as shown in figure 4(c), on one hand leads to continuation of aging for $\tilde{\sigma} < \tilde{\sigma}_y$. On the other hand, for $\tilde{\sigma} \geq \tilde{\sigma}_y$ material eventually undergoes rejuvenation producing a liquid phase. For such conditions, we plot evolution of strain ($\gamma$) under application of $\tilde{\sigma}$ for $\phi_i$ in the domain $\phi^* < \phi_i < \phi_c$ in the inset of figure 7. It can be seen that for $\tilde{\sigma} < \tilde{\sigma}_y$, $\gamma$ increases but eventually reaches a plateau. However, for $\tilde{\sigma} \geq \tilde{\sigma}_y$, $\gamma$ shows a sharp increase with time. Application of $\tilde{\sigma}$ in the vicinity of $\tilde{\sigma}_y$ but slightly larger and smaller than $\tilde{\sigma}_y$, can be seen to be following very similar evolution of $\gamma$ for a significant period of time. However, in the limit of very long times, $\gamma$ bifurcates. This phenomenon is popularly known as viscosity bifurcation in the literature. For strain curves associated with $\tilde{\sigma} \geq \tilde{\sigma}_y$, we can define the time at the point of inflation $d^2\gamma/d\tilde{t}^2 = 0$ as the time to yield ($\tilde{t}_{dy}$). In figure 7 we plot $\tilde{t}_{dy}$ as a function of $\tilde{\sigma}_y$. It can be seen that time $\tilde{t}_{dy}$ rapidly increases as $\tilde{\sigma} - \tilde{\sigma}_y \to 0$. On the other hand, for $\tilde{\sigma} >> \tilde{\sigma}_y$, $\tilde{t}_{dy}$ decreases weakly with increase in $\tilde{\sigma}$.

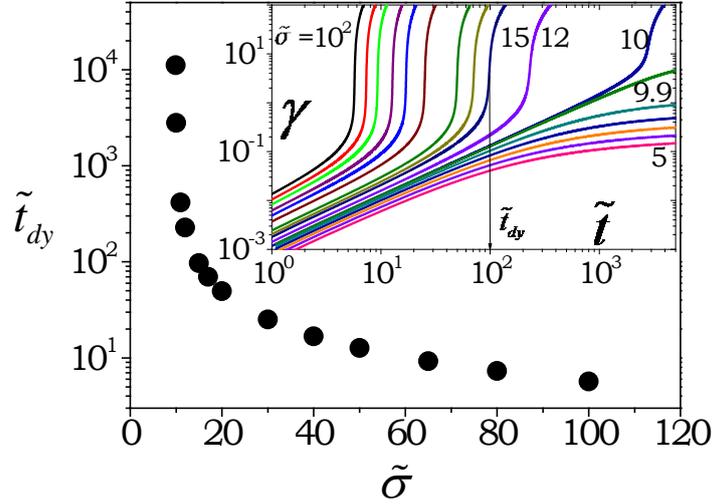

**Figure 7.** Time to yield $\tilde{t}_d$ is plotted against creep stress for $\mu=1.2$, $A=10$ (the corresponding $\phi_c=0.74$, $\tilde{\sigma}_c=9.7$ and $\phi^*=0.6$), $\phi_i=0.7$ (the corresponding $\tilde{\sigma}_{yi}=9.93$). It can be seen that as stress decreases towards yield stress, $\tilde{t}_d$ increases very sharply. The inset shows evolution of strain as a function of time for the values of stress from left to right $\tilde{\sigma}=$ 100, 80, 65, 50, 40, 30, 20, 17, 15, 12, 10, 9.9, 9, 8, 7, 6, and 5. The inset clearly shows viscosity bifurcation for stress below and above $\tilde{\sigma}_{yi}=9.93$.

In figure 8 we plot evolution of $\gamma$ at constant $\tilde{\sigma}$ but at different $\phi_i$. This plot is therefore equivalent to carrying out creep experiments at different waiting times after stopping the shear melting. It can be seen that for $\phi_i$ smaller than $\phi_{ss}$ associated with $\tilde{\sigma} = \tilde{\sigma}_{ss}(\phi_{ss})$, system is in region I of figure 4b, consequently strain eventually reaches a plateau (plateau is not apparent in figure 8 as it occurs after a very long time). However, if $\phi_i$ is larger than $\phi_{ss}$, application of $\tilde{\sigma} = \tilde{\sigma}_{ss}(\phi_{ss})$ causes yielding, wherein strain can be seen to be rapidly increasing with time.

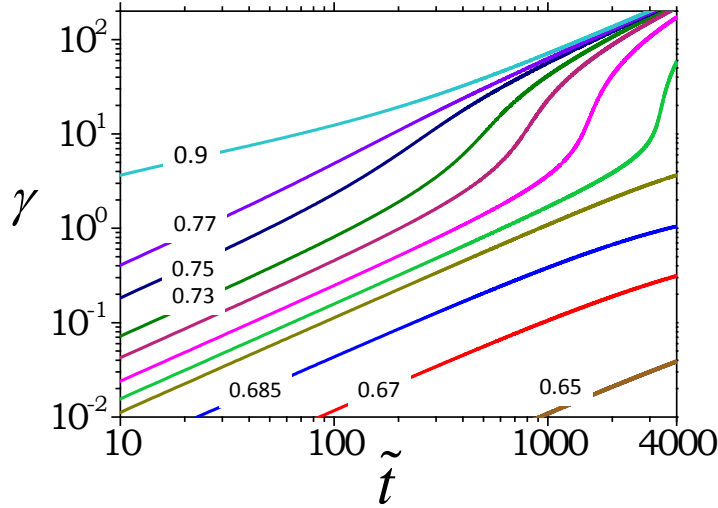

**Figure 8.** Evolution of of strain as a function of time is plotted various values of $\phi_i$ for $\mu=1.2$, $A=10$ and $\tilde{\sigma}=9.93$ (corresponding $\phi_{ss}=0.7$). From bottom to top: $\phi_i=0.65$, 0.67, 0.685, 0.698. 0.703, 0.71, 0.72, 0.73, 0.75, 0.77, and 0.9 (the corresponding $\phi_c=0.74$, $\tilde{\sigma}_c=9.7$ and $\phi^*=0.6$). It can be seen that for $\phi_i>0.7$ significant strain gets induced in the material, while for $\phi_i<0.7$, strain approaches a plateau value.

Another important characteristic feature of glassy materials in general and SGMs in specific is presence of residual stresses. Typically SGMs are shear melted by applying constant shear rate of sufficiently large magnitude prior to carrying out any rheological study. During shear melting a steady state is reached $\left(\tilde{\dot{\gamma}}_{ss} = \tilde{\dot{\gamma}}_{sm}\right)$ and the corresponding $\tilde{\sigma}_{ss}$ and $\phi_{ss}$ are given by equations (19) and (20). Subsequent to the cessation shear melting if strain is kept constant, decay in stress can be easily estimated by simultaneously solving equations (16) and (18) with $\tilde{\tau}$ given by equations (10) to (12) and initial condition of $\tilde{\sigma} = \tilde{\sigma}_{ss}$ and $\phi = \phi_i = \phi_{ss}\big|_{\tilde{\dot{\gamma}}_{ss}=\tilde{\dot{\gamma}}_{sm}}$ at $\tilde{t} = 0$, where $\tilde{\dot{\gamma}}_{sm}$ is dimensionless shear rate associated with shear melting. It should be noted that, even though strain is kept constant resulting in $\dot{\gamma}=0$, $\dot{\gamma}_E$ and $\dot{\gamma}_V$ may not be constant leading to: $\dot{\gamma}_E = -\dot{\gamma}_V$. As stress relaxes, spring in the Maxwell model contracts, giving rise to: $\dot{\gamma}_V = \tilde{\sigma}(t)/\tilde{G}$, where $\tilde{\sigma}(t)$ is an instantaneous stress remained in a material as it relaxes. In figure 9 we plot $\tilde{\sigma}$ as a function of $\tilde{t}$ for a material with constant modulus $(\tilde{G} = 1)$ with $A=10$ and $\mu=1.1$, 1 and 0.9 for various values of $\phi_i$ in the range 0.95 and 0.65. It can be seen that higher the value of $\phi_i$ is, greater is the plateau value of $\tilde{\sigma}$ in the limit of $\tilde{t} \to 0$. Furthermore this value is independent of $\mu$ as per equation (20). Figure 9 shows that for $\mu=0.9$ stress decays to 0, while for $\mu=1$ stress shows a power law decay. For $\mu=1.1$, on the other hand, stress shows a plateau in the limit of high times describing a presence of residual stress. The most prominent feature of figure 9 is that irrespective of the initial value of stress, in the limit of long times all the stress relaxation curves coincide for a given value of $\mu$. Consequently, according to the present model, the residual stress is independent of the initial stress or a state of a material.

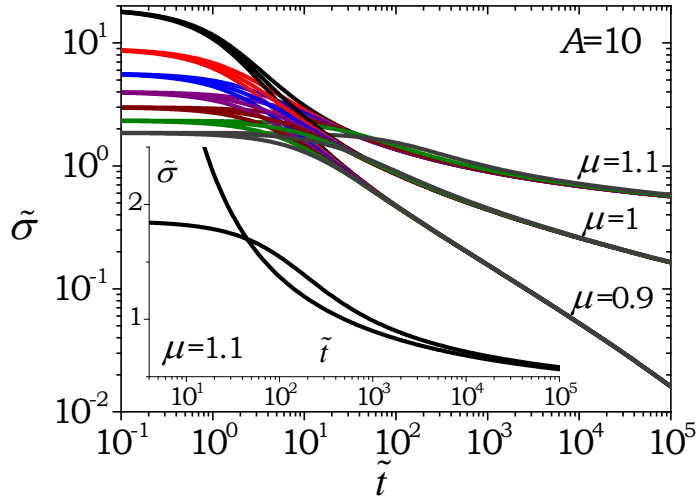

**Figure 9.** Relaxation of stress subsequent to cessation of shear melting for a material with constant modulus ($\tilde{G}=1$) for different values of shear melting shear rates $\phi_i$ (or $\dot{\gamma}_{sm}$) and $\mu$ for $A=10$. For a given value of $\mu$, $\tilde{\sigma}$ in the limit of $\tilde{t} \to 0$ only depends on $\phi_i$. In that limit, from top to bottom: $\phi_i=0.95$, 0.9, 0.85, 0.8, 0.75, 0.7, and 0.65. The corresponding $\dot{\gamma}_{sm}$ depends on $\mu$ and can be obtained from equation (19). In the limit of $\tilde{t} \to \infty$, stress shows a plateau for $\mu>1$, stress undergoes power law relaxation for $\mu=1$, while stress decays to 0 for $\mu<1$. In the inset $\tilde{\sigma}$ is plotted as a function of $\tilde{t}$ for $\mu=1.1$ for two values of $\phi_i=0.95$ and 0.65. The inset shows that greater initial stress

leads to faster relaxation of stress due to rejuvenation caused by dissipative deformation of dashpot as a result of contracting spring.

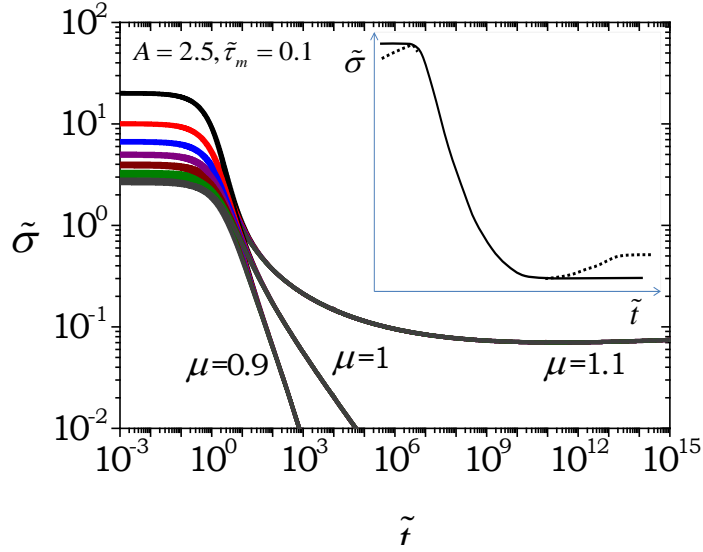

**Figure 10.** Stress is plotted as a function of time for a material with time dependent modulus given by equation (15) for various model parameters as mentioned. For a given value of $\mu$, $\tilde{\sigma}$ in the limit of $\tilde{t} \to 0$ only depends on $\phi_i$ whose values are same as that mentioned in figure 9. It can be seen that since modulus increases with time, residual stress in the material with $\mu > 1$ may show increase at very large times. However, as shown in the inset since modulus always remains finite, in the limit of long time stress must show a plateau in that limit even if it shows an increase over a certain period. The inset also shows a possibility that at very early times for $\tilde{t} << \tilde{\tau}(\tilde{t})$ possible increase in modulus may show an early increase in stress.

In addition to the relaxation time, if modulus of a material also shows an increase, relaxation of stress shows some further interesting features. It is well known that increase in modulus of a spring having constant strain increases the stress induced in the same. Consequently increase in modulus as a function of time impedes relaxation of stress. In figure 10 we plot relaxation of stress for $\mu = 1.1$, 1 and 0.9 for different values of $\phi_i$. Various features of the observed behavior are qualitatively identical to that for a material with constant modulus (shown in figure 9) for $\mu \leq 1$. This suggests that irrespective of whether modulus increases or not, stress must decay complete for $\mu \leq 1$. However for $\mu =1.1$, at large times the relaxation curves in Figure 10 are observed to demonstrate a minimum, which can be attributed to time dependent increase in modulus. Nonetheless, as mentioned before, as relaxation time diverges to $\infty$, modulus eventually must reach a constant value. Consequently residual stress also must reach a constant value. In the inset of figure 10, we represent a schematic wherein possible scenarios are described. Depending upon when modulus reaches a constant in relation with increase in relaxation time, stress may or may not show a minimum before reaching a residual stress plateau in the limit of long times. In the limit of very small times, if modulus shows enhancement, stress may also show increase in that limit before beginning to relax. Although to best of our knowledge increase in stress during stress relaxation of aging SGMs has not been reported in the literature, the present work clearly predicts such possibility particularly for those materials that show very prominent increase in modulus as a function of time.

Subsequent to cessation of shear melting, instead of keeping strain constant, if stress is removed ($\tilde{\sigma}=0$) the material will undergo strain recovery. It is known that upon removal of stress a single mode Maxwell model undergoes an instantaneous recovery.[57] However in real viscoelastic (including soft glassy) materials recovery occurs over a finite (and sometimes a prolonged) period of time. The period over which recovery takes place is controlled by retardation timescale associated with a material. Therefore, in order to solve a strain recovery problem, we consider a dashpot (with viscosity $\eta_d$) in parallel with the spring. Consequently the corresponding Voigt element (spring and dashpot in parallel) will have a retardation time given by: $\tau_d = G/\eta_d$, where $G$ is the modulus associated with the spring. It should be noted that in addition to Voigt element there also exists a dashpot with viscosity $\eta$ in series (same as that of Maxwell model), by virtue of which the system also has a relaxation time $\left(\tau = G/\eta\right)$. However this series dashpot does not play any role during recovery as the deformation of the same is always permanent, consequently $\tilde{\dot{\gamma}}_V = 0$. We assume that $\tau_d$ represents an average retardation time of a material, whose average relaxation time is ... However if relaxation time undergoes a time dependent evolution according to equation (8), causality demands that retardation time also must show the identical time dependence.[53] As a result, the mean retardation time is given by:[53]

$$\tilde{\tau}_d = \alpha \tilde{\tau}, \tag{29}$$

where $\alpha$ is a constant and $\tilde{\tau}_d = \tau_d / \tau_0$. The elastic strain recovery upon removal of stress subsequent to cessation of shear melting with initial condition: at $\tilde{t} = 0$, $\gamma = \gamma_{sm} = \tilde{\sigma}_{ss} / \tilde{G}_{ss}$ and $\phi = \phi_i = \phi_{ss}|_{\tilde{\dot{\gamma}}_{ss} = \tilde{\dot{\gamma}}_{sm}}$ is given by:

$$\ln\left(\frac{\gamma}{\gamma_{sm}}\right) = \frac{\ln(\phi_i/\phi^*)}{\alpha}\left[\left(1 + \left[\frac{\ln\phi^*}{\ln(\phi^*/\phi_i)}\right]^{\frac{1}{1-\mu}} A\tilde{t}\right)^{1-\mu} - 1\right]\cdots \mu > 1, \tag{30}$$

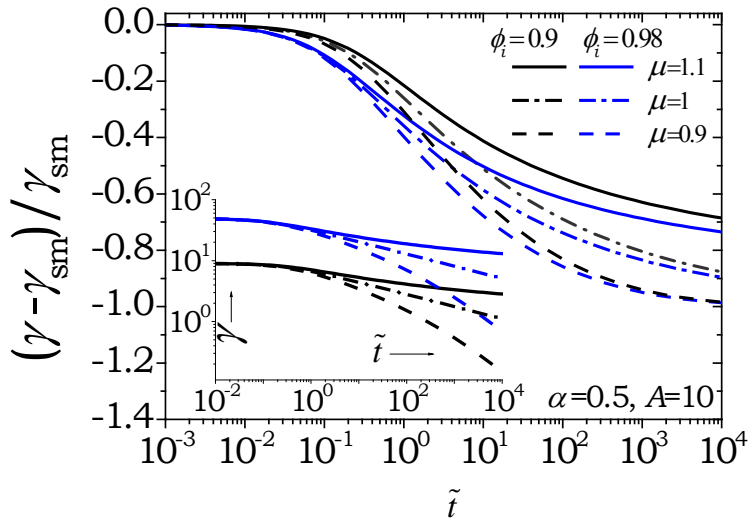

**Figure 11.** Evolution of $(\gamma - \gamma_{sm})/\gamma_{sm}$ is plotted for various values of $\mu$ and $\phi_i$. In the inset identical data is plotted for elastic strain present in a material as a function of time. It can be seen that for $\mu \leq 1$ entire elastic strain is recovered in the limit of long times, however for $\mu > 1$ residual elastic strain remains in a material.

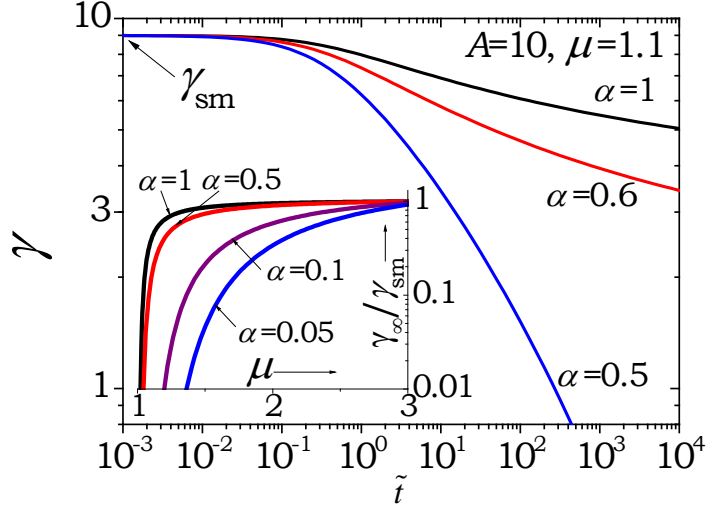

**Figure 12.** Elastic strain present in a material is plotted as a function of time for different values of $\alpha$. It can be seen that increase in $\alpha$ increases the rate at which strain is recovered. In the inset normalized ultimate elastic strain (residual strain) is plotted as a function of $\mu$, which shows that $\gamma_\infty$ increases with both, $\alpha$ as well as $\mu$.

$$\gamma = \gamma_{ss}\left(1 + \frac{A\tilde{t}}{\phi_i^{-A}}\right)^{-\frac{1}{\alpha A}} \quad \ldots \quad \mu = 1, \text{ and} \tag{31}$$

$$\ln\left(\frac{\gamma}{\gamma_{sm}}\right) = \frac{(1-\mu)A\ln\phi_i - 1}{\alpha A(1-\mu)}\left[\left[1 + \frac{A\tilde{t}}{\left(1-(1-\mu)A\ln\phi_i\right)^{1/(1-\mu)}}\right]^{1-\mu} - 1\right] \quad \ldots \quad \mu < 1$$

$$\tag{32}$$

The ultimate recovered strain $(\gamma_\infty)$ can be obtained from equations (30) to (32) in the limit of $\tilde{t} \to \infty$ and is given by:

$$\ln\left(\frac{\gamma_\infty}{\gamma_{sm}}\right) = \frac{\ln\left(\phi^*/\phi_i\right)}{\alpha} = \frac{1-(1-\mu)A\ln\phi_i}{\alpha A(1-\mu)} \quad \ldots \quad \mu > 1$$

$$\tag{33}$$

$$\gamma_\infty = 0 \quad \ldots \quad \mu \leq 1$$

In figure 11 we plot $(\gamma - \gamma_{sm})/\gamma_{sm}$ as a function of $\tilde{t}$ for three values of $\mu$ and two values of $\phi_i$ as represented by equations (30) to (32). In the inset we plot identical data in terms of time dependent recovery of $\gamma$. It can be seen that for $\mu \leq 1$ the total elastic strain $\gamma_{sm}$ indeed gets recovered in the limit of long times. However for $\mu > 1$ only part of the elastic strain gets recovered leading to presence of residual elastic strain in a material. This is because; owing to aging, average retardation time of a material diverges converting the dashpot, which is in parallel with the spring, into a rigid rod preventing any further recovery of the spring. In figure 12 we plot an effect of average retardation time by varying factor $\alpha$ on the recovery behavior. It can be seen that decrease in $\alpha$, which corresponds to decrease in retardation time at any fixed aging time, magnitude as well as the rate of recovery increases. In the inset we plot $\gamma_\infty / \gamma_{sm}$ as a function of $\mu$, which clearly shows that larger the value of $\mu$ or $\alpha$ is, smaller is the ultimate recovered strain $\left(\gamma_\infty\right)$. Interestingly in the limit of $\alpha \to 0$ all the elastic strain is expected to undergo an instantaneous recovery irrespective of the value of $\mu$.

IV.    Discussion

The most prominent result of the proposed model is that, for a material with time dependent modulus for $\mu > 1$, the steady state relation between stress and strain rate is non-monotonic. On the other hand, for a constant modulus with $\mu > 1$ a material show a plateau of constant stress in the limit of small strain rate. For while for $\mu \leq 1$ the steady state flow curve is always monotonic. We believe that this result is not limited to only power law dependence of relaxation time on waiting time. Any dependence between relaxation time and waiting time, which is stronger than linear, must show a behavior similar to that observed for $\mu > 1$. Conversely any dependence which is weaker than linear should result in monotonic dependence between steady state stress and strain rate. The non-monotonic relation between stress and strain rate for the present model gives rise to thixotropic yield stress. As described in figure 5, yield stress remains constant until $\phi_i$ remains larger than $\phi_c$, below which it shows a logarithmic dependence on time. Recently Negi and Osuji[58] measured yield stress and yield strain of 4 days old 3.5 weight % aqueous suspension of Laponite. They observed that yield stress indeed showed a constant value for a certain period of time beyond which it showed a logarithmic increase with time. Interestingly relaxation time of the studied Laponite suspension showed exponential dependence on waiting time over a same period for which constant yield stress was observed. At higher times Laponite suspension showed a power law dependence on waiting time with $\mu \approx 1.8$. The yield stress in the corresponding regime showed logarithmic increase with respect to waiting time. According to the present model it appears that for Laponite suspension studied by Negi and Osuji,[58] relaxation time followed two different dependences on $\phi$: for $\phi > \phi_c$, $\tau = \tau(\phi)$ leads to $\tau \sim \exp(t/\tau_0)$, while for $\phi < \phi_c$, $\tau = \tau(\phi)$ leads to $\tau \sim (t/\tau_m)^\mu$. Consequently, similar to that shown in figure 5, the model is indeed expected to predict constant value of yield stress for $\phi > \phi_c$ followed by a logarithmic increase. It is important to note that logarithmic increase in modulus during aging as predicted by the present model using a scaling relation, which in turn is responsible for logarithmic increase in yield stress, has been observed for many SGMs.[32, 51, 59]

Negi and Osuji[58] also observed that the yield strain decreases over the regime where yield stress is observed to be constant (for small times). On the other hand, yield strain is observed to be constant in a limit of long times when yield stress is observed to increase logarithmically. In the present model, considering the yield strain to be: $\gamma_y = \tilde{\sigma}_y / \tilde{G}$, its dependence on $\tilde{t}$ can be directly written as:

$$\gamma_y = \frac{\tilde{\sigma}_c \ln(1/\tilde{\tau}_m)}{\ln(1/\tilde{\tau}_m) + \mu \ln\left[\left[\tilde{\tau}(\phi_{sm})\right]^{1/\mu} + A\tilde{t}\right]} \quad \ldots \text{ for } \ldots \phi_i \geq \phi_c \text{ or } \tilde{t} \leq \tilde{t}\big|_{\phi=\phi_c}$$

(34)

$$\gamma_y = \frac{\phi_i}{(1-\phi_i)} = \left[\exp\left(\frac{1 - \left[\left[\tilde{\tau}(\phi_{sm})\right]^{1/\mu} + A\tilde{t}\right]^{1-\mu}}{A(\mu-1)}\right) - 1\right]^{-1}$$

$$\ldots \text{ for } \ldots \phi^* < \phi_i < \phi_c \text{ or } \tilde{t} > \tilde{t}\big|_{\phi=\phi_c}$$

(35)

Equation (34) very clearly suggests that for $\phi_i \geq \phi_c$ (small times) $\gamma_y$ should decrease with increase in $\tilde{t}$. On the other hand, for $\phi^* < \phi_i < \phi_c$, in the limit of long times the term in the braces of equation (35) tends to 1 leading to constant value of $\gamma_y$. Overall the present model explains the yielding behavior of Laponite suspension reported by Negi and Osuji[58] very well.

It is well known that any material that possesses a yield stress shows shear banding in a flow field having a gradient of shear stress. Axial flow of a yield stress fluid in a pipe is a classic textbook example of shear banding.[54] However even in absence of shear stress gradient, a material with a non-monotonic steady state relationship between $\sigma_{ss}$ and $\dot{\gamma}_{ss}$, which is observed for $\mu > 1$, demonstrates (thixotropic) shear banding if imposed shear rate is less than $\dot{\gamma}_c$ (refer to figure 4(b)). This is because negative slope of $\sigma_{ss} - \dot{\gamma}_{ss}$ dependence is constitutionally untenable, consequently $\dot{\gamma}_{ss}$ does not exist below $\dot{\gamma}_c$. Let us consider a case of simple shear flow in between parallel plates separated by distance $H$. If the top plate velocity $V$ is such that $V/H < \dot{\gamma}_c$, shear banding will take place so that band (or bands) having (total) thickness $h = V/\dot{\gamma}_c$ will flow with $\dot{\gamma}_c$. On the other hand, a band (or bands) with total thickness $H - h$ will remain stationary. Increase in $V$ will decrease a width of the stationary band(s) and in the limit of $V/H = \dot{\gamma}_c$, entire sample will flow with shear rate $\dot{\gamma}_c$. The present model very clearly suggests that the thixotropic shear banding is possible only when $\mu > 1$ and $\tilde{G}$ increases sufficiently strongly so that solution of equation (22) makes $\phi_c$ to lie in the range: $\phi^* < \phi_c < 1$. Remarkably it is indeed observed that simple concentrated emulsion which shows negligible enhancement in modulus does not show thixotropic shear banding, but clay loaded emulsion which shows significant enhancement in modulus does show thixotropic shear banding[13] as suggested by the present model. Interestingly Bécu et al.[60] suggested that in a simple concentrated emulsion if attractive interactions are induced, it shows thixotropic shear banding. Although Bécu et al.[60] do not measure the modulus, we believe that attractive interactions will indeed induce evolution of modulus in accordance with the present model. Experimentally such behavior has also been observed for variety of SGMs such as suspensions of charged particles including smectite clay,[61-63] cement paste,[63] for which not just $\mu$ is expected to be greater than unity but modulus also shows prominent increase as a function of time. The present work therefore also suggests that polymeric materials undergoing crosslinking reaction, wherein relaxation time shows stronger that linear dependence on time and modulus shows prominent increase,[53] should also demonstrate shear banding.

The very fact that the steady state relation between stress and strain rate is monotonic for $\mu \leq 1$ implies absence of thixotropic yield stress. Consequently, a material with $\mu \leq 1$ must yield for any value of applied stress. However, as apparent from figure 6(a), even with $\mu \leq 1$, smaller the stress is larger time it takes to stop enhancement of relaxation time. In practice yield stress is estimated by applying linear or oscillatory stress ramp. Since stress increases from a small value to a large value over a finite time, at a certain stress material shows sudden enhancement in strain. As a result material shows apparent yield stress, which is greater than zero. This behavior, therefore, may manifest itself as undergoing weak flow below a certain stress and strong flow above certain stress, thereby resulting in so called "engineering yield stress." Furthermore engineering yield stress is expected to decrease with decrease in the rate at which stress is increased. Presence of such engineering yield stress has indeed been reported by Derec et al.[51] for a moderate concentration (36 to 44 volume %) suspension of 100 nm silica particles with $\mu = 0.55$.

Application of stress also affects the rate of evolution of relaxation time ($\mu_t$). In the literature, $\mu_t$ has been experimentally estimated as a function of stress for soft microgel paste[24] and aqueous suspension of Laponite.[27] It has been observed that in the limit of small stresses $\mu_t \to \mu$, while in the limit of large stresses $\mu_t \to 0$. As shown in figure 6, the model predicts this behavior very well. Figure 6 also shows negative values of $\mu_t$. Experimentally it is indeed observed that application of stress not just decreases the rate of change of relaxation time but also the relaxation time itself, thereby justifying presence of negative values of $\mu_t$ as predicted by the model.

Viscosity bifurcation has been observed for many SGMs such as Laponite suspension, bentonite suspension, mustard, hair gel, mayonnaise, foam, quick sand (mixture of fine sand, clay and salt water), physical gel with polymeric backbone, etc.[9, 10, 32, 64, 65] While for some of these materials the value of power law exponent $\mu$ is not reported, for others it is around or above 1. Strictly speaking the present model predicts viscosity bifurcation for $\mu > 1$. However, time taken by the material to undergo substantial or noticeable flow is very large. Consequently even for $\mu$ less than but close to 1 effect of viscosity bifurcation can be observed experimentally.

Another rheological behavior closely related to viscosity bifurcation is delayed yielding, which can occur for two cases. For $\mu \leq 1$ smaller the stress is, delayed will be the strain induced in a material (apparent yielding). On the other hand, for $\mu > 1$ yielding will get delayed as yield stress is approached from higher values as shown in figure 7. Sprakel and coworkers [16] studied thermo-reversible stearylated silica gels, and weak depleted gel of polystyrene particles and observed delayed yielding no matter how small the stress is. Although Sprakel and coworkers[16] do not measure value of $\mu$, since yielding is observed for all the studied stresses, it could be possible that it is below 1. Sprakel also observe that with decreases in stress, time to yield increases faster at small stresses while slower at large stresses. Interestingly figure 7 qualitatively captures this behavior. Baldewa and Joshi[15] also observed delayed yielding for around 80 days old aqueous Laponite suspension for which $\mu$ under quiescent conditions is observed to be slightly below 1 in agreement with the present model.

In the present model we employ only a single mode, and competition between aging and rejuvenation of the same respectively leads to decrease and increase in frsuppee energy. As a result, all those rheological effects for which consideration of only a single mode is sufficient can be explained by the model proposed in this work. On the other hand, there are many other important effects that depend strongly on how shape of relaxation time spectrum is affected by competition between aging and rejuvenating modes. Consequently, effects such as viscosity bifurcation, presence of engineering yield stress, shear banding, which can in principle be explained by a single mode model, get strongly influenced by dynamically changing relaxation time spectrum. Many SGMs have also been observed to show overaging,[20, 21] wherein application of moderate magnitude of deformation field increases the relaxation time rather than decreasing it. This effect has also been attributed to alteration of relaxation time distribution.[19]

It is known that perfectly crystalline materials (or perfect solids) do not relax over any timescale. Consequently, upon application of step strain, stress induced in the same remains unrelaxed for an indefinite period of time. It is therefore no surprise that the glassy materials including soft glasses, which are in apparent solid state, cannot relax the induced stress completely over the practically measurable time scales. Very recently, Ballauff and coworkers[22] studied stress relaxation subsequent to shear melting by using MCT and molecular dynamics simulations as well as by carrying out experiments on two types of SGMs: particulate colloidal glasses with hard sphere interactions and PS-PNiPAM core shell suspension. They observed that below a certain threshold volume fraction (or above a temperature for MD simulations), stress decays completely while at high volume fractions the materials indeed demonstrates presence of residual stresses. They observed that the volume fractions for which the residual stress is observed, stress relaxes by about a factor of ten or less before plateauing out. Importantly, MCT, which does not account for aging, while shows residual stress above a certain concentration, the stress does not relax at all before showing a plateau, thereby showing a partial disagreement with the experimental data.

Such residual stress can originate from two factors. It is possible that immediately after shear melting is stopped the particles get arrested in such a fashion that faster modes associated with smaller length-scales are finite but slower modes associated with larger length-scales are practically infinite.

However there is no time dependent evolution of the relaxation modes. Under such case a material relaxes only up to such an extent allowed by finite modes. The other possibility is that immediately after cessation of shear melting all the timescales are finite, which age as a function of time. Eventual divergence of such relaxation timescales over finite time does not allow complete relaxation of stress.

The present model can, in principle, represent both the possibilities, however the match is qualitative since the model is limited by a single mode. The present model can express the first possibility by considering $\mu >> 1$, wherein relaxation time diverges soon after shear melting is stopped. However, in this case owing to consideration of only a single mode, relaxation of stress will not be very significant as is the case with MCT. However, since relaxation modulus is given by: $G(t) = \Sigma G_i e^{-t/\tau_i}$, consideration additional finite relaxation modes may represent the decay of stress before it plateaus out. The second case is represented in figure 11, wherein single mode with $\mu > 1$ can be seen to predict the right magnitude of decay. Furthermore Ballauff and coworkers[22] observe that greater the stress (or shear rate) induced during shear melting, faster is the relaxation of stress. In the inset of figure 11, we plot two relaxation curves subsequent to shear melting at different rejuvenation stresses (or shear rates). The model indeed predicts that the relaxation is faster when shear melting stress is higher. This because higher shear stress at the time of cessation of shear melting induces to greater $\tilde{\dot{\gamma}}_V$ in the dashpot (in opposite direction), which causes partial rejuvenation leading to slower increase in relaxation time. This facilitates greater relaxation of stress at early times as shown in in the inset of figure 11. However in the limit of long times all the relaxation curves, irrespective of shear melting stress/strain rate for a given $\mu$, superpose. Consequently, the present model shows that residual stress (or stress in the limit of very large times) is independent of the applied shear melting shear rate. The experiments of Ballauff and coworkers[22] show that residual stress shows weak increase with increase in shear melting shear rate. While those systems wherein stress decays completely, stress in the limit of very large but at identical time shows decrease with decrease in shear melting shear rate. We believe that this difference in the model prediction and the experimental results is due to consideration of only a single mode.

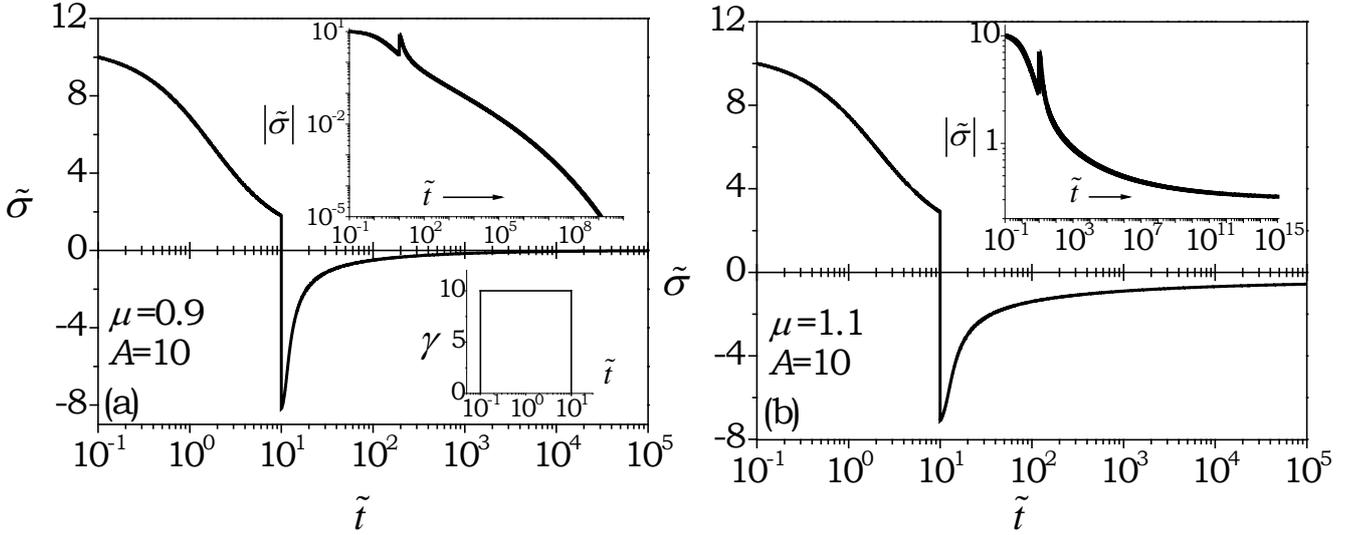

**Figure 13.** Stress response to switch on and off strain profile shown in the lower inset of figure (a) for a material having $\tilde{G}=1$. For $\mu=0.9$, $\tilde{\sigma}\to 0$ in the limit $\tilde{t}\to\infty$ as shown in part (a), while for $\mu=1.1$, $|\tilde{\sigma}|>0$ in the limit $\tilde{t}\to\infty$. In both the figures the top inset describes $|\tilde{\sigma}|$ plotted on a logarithmic scale. This figure therefore suggests that for $\mu\leq 1$ a material shows weak long term memory, while for $\mu>1$ a material shows strong long term memory.

Based on the stress relaxation behavior Fielding and coworkers proposed a distinguishing criterion of weak and strong long term memory for SGMs. They suggested an experiment wherein a material is subjected to step strain at time $t_0$, which is switched off at time $t_1$, and the relaxation of stress is monitored for $t>t_1$. According to their proposal if $\sigma\to 0$ in the limit of $t\to\infty$, it has weak long term memory. On the other hand, in that limit if finite residual stress remains in the material it has strong long term memory. We solve the present model (with constant modulus) for the suggested experiment by subjecting it to a suggested flow field shown in the lower inset of figure 13(a). The model prediction for the two cases $\mu < 1$ and $\mu > 1$ is shown in figure 13(a) and (b) respectively. The model clearly predicts that materials with $\mu \leq 1$ have a weak long term memory while materials with $\mu > 1$ have strong long term memory.

The strain recovery behavior of many SGMs such as microgel paste,[24] aging surfactant paste,[23] mustard,[32] clay suspension,[32] colloidal gel,[32] etc., has also been studied in the literature. The qualitative nature of the strain recovery in these systems is similar to that described in figure 11. The model also predicts presence of residual strain for materials with $\mu > 1$. However the experiments cannot report residual strain as it is difficult to distinguish between residual strain and irrecoverable strain due to flow (dissipation). In polymeric glasses, residual strains are known to cause distortion (warpage) of the end product.[66, 67] Usually soft glassy commercial products are in macroscopically unstressed state, however presence of residual strain may lead to local pockets of residual stress, which may adversely affect the long time behavior of the materials.

The results of the proposed model, though it uses only a single mode, render insight into how variation in relaxation time (represented by $\mu$ and $A$) and modulus (represented by $\tilde{\tau}_m$) affect various rheological behaviors. Among these parameters, value of $\mu$, which represents $d\ln\tau/d\ln t$, is primarily responsible for determining the material behavior. Firstly $\mu = 0$ represents material in equilibrium state that does not undergo any evolution as a function of time. If $\mu \leq 1$, the model shows that the steady state stress – strain rate relationship (flow curve) is monotonically increasing. Consequently a material flows at all the stresses, and therefore does not demonstrate presence of true yield stress. However owing to time dependency material does demonstrate thixotropy. Furthermore, as $\mu$ approaches unity from below, material may show 'engineering yield stress' or 'apparent delayed yielding' depending upon experimental conditions. For $\mu > 1$, qualitative behavior of flow curve is different depending upon how modulus scales with time. For the materials whose aging dynamics is purely entropic modulus remains constant during aging. Under such conditions (constant modulus and $\mu > 1$) flow curve is monotonic but plateaus out as strain rate decreases. In this case material shows thixotropy as well as true yield, which is independent of time. If the inter-particle energetic interactions affect the aging behavior, modulus increases as a function of time. Although scaling relation derived in this work suggests modulus to follow equation (15), a nature of the flow curve can be predicted for any functional form: $\tilde{G} = \tilde{G}(\tilde{t})$. The key is location of $\phi_c$ given by equation (22) with respect to location of $\phi^*$ given by equation (13). If $\phi^* > \phi_c$, flow curve would be qualitatively similar to that for a system with constant modulus. However if $\phi^* < \phi_c$ flow curve will be non-monotonic as shown in figure 4(b) leading to time dependent (thixotropic) yield stress along with thixotropy. A limit of $\mu >> 1$ represents extremely fast evolution of relaxation time as a function of time. Consequently relaxation time diverges very rapidly freezing the system kinetically in a high free energy state. An interesting example of such limit of $\mu >> 1$ is a system of dense granular materials. In this system subsequent to rejuvenation particles get arrested in random close packing configuration which is a high free energy state. The limit of $\mu >> 1$ is also observed during physical or chemical gelation, wherein owing to bond formation mobility of the constituents rapidly decreases causing divergence of relaxation time. Furthermore, even though the present model cannot predict the behaviors such as delayed yielding with a minimum in strain rate as observed by Sprakel et al.[16] and delayed solidification, it is expected that increase in $\mu$ would enhance possibility of delayed solidification while decrease in $\mu$ would enhance possibility of eventual yielding.

There are important differences between the present model compared to the other models such as fluidity/thixotropic, MCT and SGR. Firstly the primary framework of the present model is evolution of free energy. Consequently a material response gets divided into two regimes. In the first one, a material eventually acquires the equilibrium state ($\mu \leq 1$) and in the other it does not ($\mu > 1$). Importantly this demarcation is physically intuitive and the parameter $\mu$ can be experimentally obtainable. In various fluidity/thixotropic models dependence of viscosity on a structure parameter $\lambda$ is arbitrarily assumed so as to demonstrate various rheological effects including non-monotonic steady state flow curve. The present model, on the other hand, proposes a relation between relaxation time and free energy, which shows an experimentally observed time dependence of relaxation time that in turn shows various rheological effects as discussed. Very importantly, to best of our knowledge, the present model is the only model that accounts for time dependence of modulus. Moreover we actually attribute the non-monotonicity of the steady state flow curve leading to various thixotropic effects to the time dependency of the modulus as vindicated by experiments on many different kinds of SGMs. Consequently, a material behavior, in principle can be a priory guessed simply based on the behavior of relaxation time and modulus, which in our opinion is the most prominent feature of the present model.

The models such as MCT and SGR, on the other hand, are mathematically involved, however give greater insight into the glassy dynamics. Out of these models, MCT does not involve aging dynamics, and consequently either shows a glass state or a liquid state based on the concentration. As a result stress in the glass state does not relax at all as shown by Ballauff and coworkers,[22] contrary to experimental behavior, which shows relaxation before plateauing out. While SGR model is primarily based on aging dynamics, rejuvenation is induced by strain. Consequently application of finite strain rate causes complete rejuvenation in the SGR model. The present framework on the other hand considers rejuvenation in terms of strain rate and complete rejuvenation, therefore is possible only in the limit of infinite strain rate. Furthermore, SGR model considers only a full aging scenario ($\mu = 1$), unlike the present model, that considers $\mu$ as a parameter. Consequently SGR model does not predict residual stress at all, which also is the case with the present model for $\mu = 1$. In addition SGR model also does not predict various effects arising from time dependent modulus. The most significant feature of the SGR model is the rigor involved in the analysis which leads to consideration of relaxation time spectrum and realistic prediction of alteration of the same under application of various kinds of deformation fields. The present model is based on simple first order kinetics leading to evolution of a single mode relaxation time, whose effect along with time dependent modulus is considered through Maxwell model. We feel that these features of the model are an advantage, as it clearly indicates those rheological behaviors for which consideration of the first order kinetics and a single mode are sufficient.

V.   **Conclusion**

SGMs are thermodynamically out of equilibrium materials. Consequently they undergo aging wherein microstructure progressively relaxes to attain low free energy structures as a function of time. During rejuvenation, on the other hand, application of deformation field either slows down or reverses the structural recovery. The rheological behavior of SGMs therefore strongly depends on competition between aging and rejuvenation, which is responsible for many fascinating effects. In this work we present a model that considers rate of change in free energy to be a first order process and is equated to sum of decreasing (aging) and increasing (rejuvenation) contributions. Aging contribution is assumed to be proportional to excess free energy divided by timescale associated with structural rearrangement or the relaxation time ($\tau$). Consequently at smaller $\tau$, due to greater mobility of the constituents structural recovery is faster and vice a versa. The rejuvenation term is considered to be proportional to viscous component (dissipative) of rate of applied deformation field. We propose a dependence of $\tau$ on free energy, which has same functional form that proposed by Krieger - Dougherty equation or mode coupling theory in particulate suspensions. Remarkably the proposed relation leads to a power law dependence of $\tau$ on time with exponent $\mu$ in absence of any external deformation field as observed experimentally for a variety of glassy materials. We consider two cases for modulus. In the first case we consider modulus to be constant as observed for entropic aging systems. In the second case, we derive an expression for time dependence of modulus based on simple scaling arguments. Availability of relaxation time and modulus scale naturally leads to consideration of the single mode Maxwell model as a constitutive relation. The model has respectively two and three parameters depending upon whether modulus remains constant or not. All the three parameters can be estimated experimentally.

Interestingly, for $\mu > 1$, it is observed that steady state relationship between stress and strain rate is monotonic with low shear rate stress plateau when modulus is constant, while non-monotonic for time dependent modulus. The former scenario leads to thixotropy with true but constant yield stress. On the other hand, non-monotonic relation implies presence of a thixotropic (time dependent) yield stress as well as shear banding. Irrespective of the nature of modulus, for $\mu > 1$, the model predicts presence of a residual stress as well as strain. For $0 < \mu \leq 1$, on the other hand, material is observed to be merely thixotropic without thixotropic yield stress. Interestingly model also predicts decrease in $\mu$ with increase in applied stress at any given time as observed experimentally, and how $\mu$ evolves under application of stress. Finally and importantly the present model allows distinguishing between various kinds of thixotropic behaviors based on different combination of model parameters.

**Acknowledgement:** This work is supported by the department of atomic energy – science research council (DAE-SRC), Government of India.